\documentclass[manuscript]{aastex}

\newcommand{\rxte}{\textit{RXTE}}

\newcommand{\spitzer}{\textit{Spitzer}}
\newcommand\iontiny[2]{#1$\;${\tiny \scshape{#2}}}
\def\cygx1{Cygnus~X-1}
\def\chis2{$\chi^2$}
\def\msun{$M_{\odot}$}
\def\rsun{$R_{\odot}$}

\def\msunyr{$M_{\odot}$~yr$^{-1}$}

\def\ergcms{erg~cm$^{-2}$~s$^{-1}$}
\def\wm2{W~m$^{-2}$}
\def\mic{$\mu$m}
\def\cm2{cm$^{-2}$}
\def\se1{s$^{-1}$}

\def\Ave{A_{\rm V}}

\usepackage{multirow}

\shorttitle{The mid-infrared compact jets of \cygx1}
\shortauthors{Rahoui et al.}

\begin{document}


\title{A multiwavelength study of \cygx1: the first mid-infrared spectroscopic detection of compact jets}


\author{Farid Rahoui\altaffilmark{1}}
\author{Julia C. Lee\altaffilmark{1}}
\author{Sebastian Heinz\altaffilmark{2}}
\author{Dean C. Hines\altaffilmark{3}}
\author{Katja Pottschmidt\altaffilmark{4}}
\author{J\"orn Wilms\altaffilmark{5}}
\author{Victoria Grinberg\altaffilmark{5}}


\altaffiltext{1}{Harvard University, Department of Astronomy \& Harvard-Smithonian Center for Astrophysics, 60 Garden street, Cambridge, MA 02138, USA; frahoui@cfa.harvard.edu, jclee@cfa.harvard.edu}
\altaffiltext{2}{Astronomy Department, University of Wisconsin-Madison, 475. N. Charter St., Madison, WI 53706, USA; heinzs@astro.wisc.edu}
\altaffiltext{3}{Space Telescope Science Institute, N406, 3700 San Martin Drive, Baltimore, MD 21218, USA; hines@stsci.edu}
\altaffiltext{4}{CRESST, UMBC, and NASA Goddard Space Flight Center, Greenbelt, MD 20771, USA; katja@milkyway.gsfc.nasa.gov}
\altaffiltext{5}{Dr. Karl Remeis-Observatory \& ECAP, University of Erlangen-Nuremberg, Sternwartstr. 7, 96049 Bamberg, Germany; joern.wilms@sternwarte.uni-erlangen.de, victoria.grinberg@sternwarte.uni-erlangen.de}

\begin{abstract}

We report on a \spitzer/IRS (mid-infrared), \rxte/PCA+HEXTE (X-ray), and Ryle (radio) simultaneous multiwavelength study of the microquasar \cygx1, which aimed at an investigation of the origin of its mid-infrared emission. Compact jets were present in two out of three observations, and we show that they strongly contribute to the mid-infrared continuum. 
During the first  observation, we detect the spectral break $-$ where the transition from the optically thick to the optically thin regime takes place $-$ at about $2.9\times10^{13}$~Hz. We then show that the jet's optically thin synchrotron emission accounts for the \cygx1's emission beyond 400~keV, although it cannot alone explain its $3-200$~keV continuum.  
A compact jet was also present during the second observation,  but we do not detect the break, since it has likely shifted to higher frequencies. In contrast, the compact jet was absent during the last observation, and we show that the $5-30$~\mic\ mid-infrared continuum of \cygx1 stems from the blue supergiant companion star HD~226868. Indeed, the emission can then be understood as the combination of the photospheric Raleigh-Jeans tail and the bremsstrahlung from the expanding stellar wind. Moreover, the stellar wind is found to be clumpy, with a filling factor $f_\infty\approx0.09-0.10$. Its bremsstrahlung emission is likely anti-correlated to the soft X-ray emission, suggesting an anti-correlation between the mass-loss and mass-accretion rates. Nevertheless, we do not detect any mid-infared spectroscopic evidence of interaction between the jets and the \cygx1's environment and/or companion star's stellar wind.

\end{abstract}


\section{Introduction}

Accretion$-$ejection phenomena are responsible for the strong radio/infrared/X-rays correlations observed in microquasars \citep[see e.g.][]{1998Mirabel, 2002Corbel, 2003Gallo, 2009Coriat}. Indeed, some of the accreted material is returned to the interstellar medium (ISM) through collimated relativistic ejections, whose power law-like synchrotron radiation ($F_\nu\propto\nu^\alpha$) is thought to extend from the radio to the X-ray domains. The best known are the so-called compact jets, only detected in the hard state (HS), whose spectral continuum is well-modeled by a flat or weakly-inverted power law in the optically thick regime ($0\le \alpha \le 0.2$) and decreasing in the optically thin one, with $\alpha$ ranging between $-0.4$ and $-1$ depending on the electron energy distribution \citep[for a $-0.6$ canonical value,][]{1979Blandford, 1995Falcke}. To date, the spectral break where this transition takes place has only been detected photometrically in two sources, GX~339$-$4 and 4U~0614+091 \citep{2002Corbel, 2006Migliari, 2010Migliari}. Yet the knowledge of this cut-off frequency is crucial for a full understanding of the properties of the compact jets, as it is closely related to the original physical conditions, such as the magnetic field, the base radius of the jet, and the total energy of the electron population.

Among microquasars, \cygx1\ \citep{1965Bowyer} is one of the few classified as a high mass X-ray binary (HMXB). It is composed of an 8$-$16~\msun\ black hole \citep{2007Shaposhnikov, 2009Caballero} bound to the 17$-$31~\msun\ O9.7~ab supergiant HD~226868 \citep{1973Walborn, 2008Iorio, 2009Caballero}. It has a quasi-circular 5.6~d orbit \citep[][and reference therein]{1999Brocksoppb}, with an associated inclination in the range of $23^\circ-43^\circ$ \citep{2003Gies,2009Caballero}. The distance to the source is still a matter of debate, but several studies have found values between 1.5 and 2.5~kpc \citep[][and reference therein]{2005Ziolkowski, 2009Caballero}. Recently, \citet{2011Xiang} found 1.80~kpc with a 7\% accuracy from X-ray dust scattering halo studies. Moreover, \citet{2011Reid} derived 1.86~kpc with a 6\% accuracy, from radio parallax measurements obtained with the very large baseline array (VLBA) at 8.4 GHz between 2009 Jan 23 and 2010 Jan 25. Both results are consistent with 2~kpc within the uncertainties, and we adopt this value in this study.

\cygx1\ spends two-thirds of its time in the HS \citep{2006Wilms}. Indeed, although the companion is close to filling its Roche lobe, the transfer of material takes place through the strong stellar winds that are focused towards the black hole \citep[see][]{1986Giesb, 2003Gies, 2009Hanke}. This lower accretion rate transfer is less effective than for Roche lobe overflow microquasars, thereby most of the time preventing the transition to a disk-dominated soft state \citep[SS, see e.g.][]{2000Pottschmidt,2003Pottschmidt}.  Instead, the source often exhibits a so-called failed-state transition, characterized by a continuous increase of the X-ray flux up to an intermediate state (IS), followed by a strong decrease and a return to the HS, with possible launches of discrete ejections. Finally, its emission appears to be in phase with the orbital period in most of the spectral domains in which the source has been detected, \citep[][]{1999Pooley, 1999Brocksoppb, 2006Lachowicz}. This includes the radio activity, which is similar to that of other microquasars. In particular, the compact jets, characterized by a very flat spectrum ($\alpha \le 0.1$) and a 10$-$25~mJy flux density level are detected in the HS \citep{2000Fenderb, 2001Stirling}. 
\newline

In this paper, we report on a mid-infrared (MIR), radio, and X-ray multiwavelength study of \cygx1, with a focus on MIR spectroscopy. Our goals are to (1) assess the origin of the MIR emission, (2) compare the possible variation of the MIR continuum to that of the X-ray and/or radio emission, and (3) detect any spectroscopic feature arising from an interaction between the jets, the stellar wind, and/or the interstellar medium (ISM). The observations and the data reduction are presented in Sect.~2, while Sect.~3 is devoted to the X-ray and radio study. In Sect.~4 and Sect.~5,  we analyze the broad band MIR to radio SEDs of the source and the spectroscopic contents of the MIR spectra, respectively. Finally, we discuss the outcomes in Sect.~6. A companion paper, Lee et al. (in prep), will focus on the dust content of \cygx1.

\section{Observations and data reduction}

\cygx1\ was observed on UT 2005 May 23, June 7, and July 2 (P.I. S. Heinz) with the InfraRed Spectrograph \citep[IRS;][]{2004Houck} on-board the Spitzer Space Telescope \citep[\spitzer;][]{2004Werner}. In the X-ray domain, the source was observed simultaneously with the Proportional Counter Array \citep[PCA;][]{2005Jahoda} and the High Energy X-ray Timing Experiment \citep[HEXTE;][]{1998Rothschild} instruments (P.I. J. Wilms), and we made use of the 1.2$-$12~keV public data provided by the All-Sky Monitor instrument \citep[ASM;][]{1996Levine}, all mounted on the Rossi X-ray Timing Explorer $($\rxte $)$. Finally, \cygx1\ was regularly monitored at 15~GHz with the Ryle radio telescope (P.I. G. G. Pooley) and we used the 20~s and 5~min light curves that are simultaneous with the IRS spectra, for the analysis presented in this paper (see Table~\ref{logobs}).

\subsection{IRS observations}

We used the SL2 ($5.20-7.70$~\mic), SL1 ($7.40-14.50$~\mic), LL2 ($14.00-21.30$~\mic), and 
LL1 ($19.50-38.00$~\mic) modules at two nodding positions for sky subtraction. The total exposure times were each time set to 60~s in SL1 and SL2, 120~s in LL2, and 2$\times$120~s in LL1. Basic Calibration Data (BCD) were reduced following the standard 
procedure given in the IRS Data Handbook\footnote{http://ssc.spitzer.caltech.edu/irs/irsinstrumenthandbook/IRS\_Instrument\_Handbook.pdf}. 
The basic steps include bad pixel correction with {\tt IRSCLEAN} v1.9, sky subtraction, as well as 
extraction and calibration (wavelength and flux) of the spectra $-$ with the \textit{Spectroscopic Modeling Analysis and Reduction Tool } 
software {\tt SMART} v8.1.2 $-$ for each nodding position. Note that in LL2 and LL1, where the sky appeared weakly non-uniform between 
both nods, the subtraction residuals were fitted by a first order polynomial. Spectra were then nod-averaged to improve the 
signal-to-noise ratio (SNR). 

The flux calibration appeared consistent between the different modules at Obs.~1 and Obs.~2, with differences less than 5\%, and this allowed us to combine all the spectra by scaling them to SL1, which has the highest flux level with respect to SL2 and LL2. However, at Obs.~3, SL2 had a flux level about 14\% lower than that of SL1, while no such problem was detected with LL2 nor LL1. We were not able to definitively assess the reason for such a discrepancy, but a misalignement of the SL2 slit is the most reasonable explanation. We therefore scaled the SL2 flux level to that of SL1.

\subsection{\rxte\ observations}

In the reduction of the data from the two pointed instruments PCA and HEXTE, we follow the
procedure used in previous studies of Cygnus X-1
\citep[e.g.][]{2006Wilms}. Data were extracted using HEASOFT~6.9 for
all times at least 10\,minutes after passages of the South Atlantic
Anomaly. During the observations, two (first two \textsl{RXTE}
observations) and three (third \textsl{RXTE} pointing) proportional
counter units were in operation. 

\subsection{MIR data dereddening}

Our \cygx1\ observations show a strong absorption feature around 9.7~\mic. At 
the distance to the source ($\approx2$~kpc), we expect a considerable column of interstellar material, 
so this feature is most likely from silicate dust in the ISM along the line-of-sight.  Several consistent measurements of the visible 
extinction are found in the literature \citep[see][and reference therein]{2009Caballero}. In the companion paper 
(Lee et al., in prep), we derive an absolute visual extinction $\Ave\approx2.95$ from the opacity of the ISM silicate 
absorption feature $\tau_{9.7}$, defined by $\Ave=18.5\times\tau_{9.7}$ \citep{2003Draine}, consistent with that of \citet{1982Wu}, $\Ave=2.95\pm0.22$ with $R_{\rm V}=3.1$.

We therefore use this value to deredden the spectra with the extinction laws given in \citet{2006Chiar}, one of the most recent works covering the entire IRS spectral range. In their paper, the authors derived the $A_{\rm \lambda}/A_{\rm K}$ ratio rather than the usual $A_{\rm \lambda}/A_{\rm V}$ one. To express the extinction ratio in the standard way, we assign to $A_{\rm K}$ the value derived from the law given in \citet{1999Fitzpatrick} for the diffuse ISM ($R_{\rm V}\,=\,3.1$), which is $A_{\rm K}\,=\,0.111\times A_{\rm V}$. Moreover, these authors find that beyond 8~\mic, where silicate absorption dominates, the MIR extinction is different for the diffuse ISM and the Galactic center. They consequently propose two extinction laws, and we explored both options for dereddening our spectra of \cygx1. We find that the Galactic center case clearly overestimates the silicate absorption depth,  and we therefore adopt the corrections based on the extinction due to the diffuse ISM. We note that a single extinction correction 
(i.e., $\Ave=2.95\pm0.22$) produces equally good results for all three epochs.  This gives us further confidence that the observed 
silicate absorption features are caused by dust in the diffuse ISM as opposed to 
material directly associated with the \cygx1\ system.

\begin{figure}
\begin{center}
\includegraphics[width=15cm]{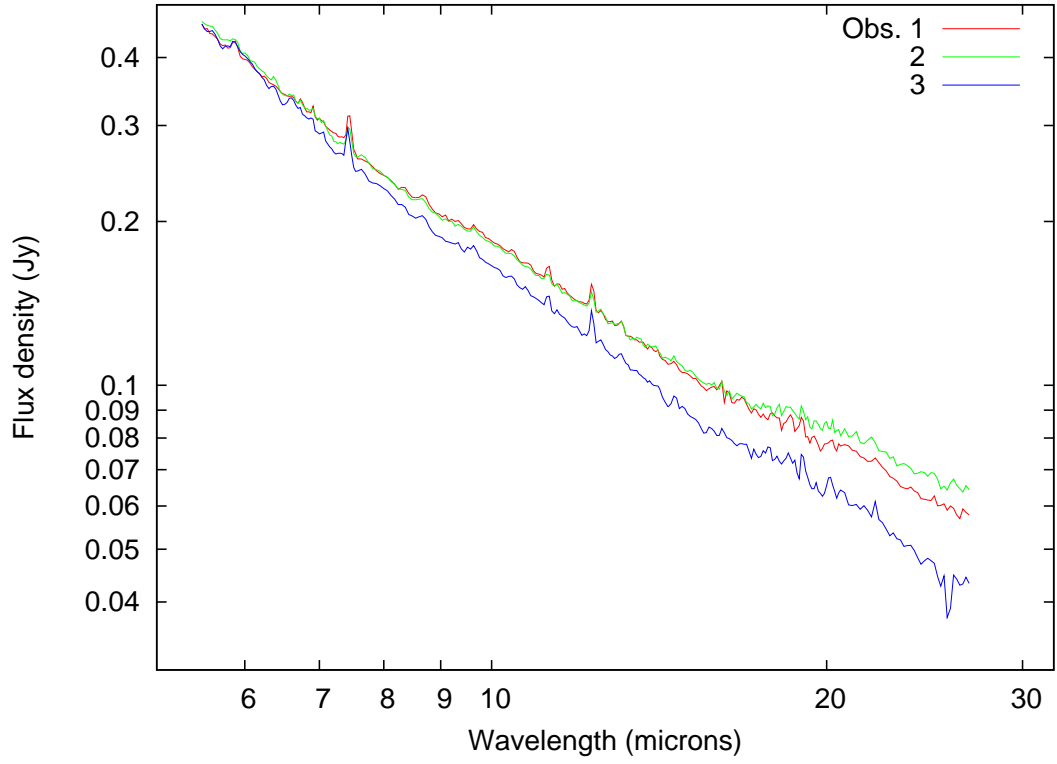}
\caption{\small Dereddened spectra of \cygx1\ based on \spitzer/IRS spectroscopy taken in 2005 May 23 (Obs.~1), June 7 (Obs.~2), and 
July 2 (Obs.~3).}
\label{specdered}
\end{center}
\end{figure}

The resulting dereddened spectra displayed in Fig.~\ref{specdered} show unambiguously that the MIR emission of \cygx1\ is variable. Indeed, the MIR continuum was brighter during Obs.~1 and Obs.~2, the difference with respect to Obs.~3 increasing with wavelength, from no difference at 6~\mic\ to an excess of 15$-$20~mJy beyond 10~\mic. 

\section{X-ray/radio behavior}

Since dust heating as a contribution to the MIR emission can be ruled out due to the lack of observed characteristic dust spectral features (Lee et a., in prep), we consider four main alternative physical processes: (1) the stellar photosphere, whose emission is constant, (2) bremsstrahlung from the expanding stellar wind, (3) synchrotron emission from jets, (4) bremsstrahlung from the accretion disk wind. We use the simultaneous radio and X-ray data to assess the extent to which the last two components are present during the observations (see Sect. 4 for details). 

\subsection{\rxte\ spectra}

\begin{figure*}
\begin{center}
\begin{tabular}{c}
Obs.~1\\
\includegraphics[height=6cm]{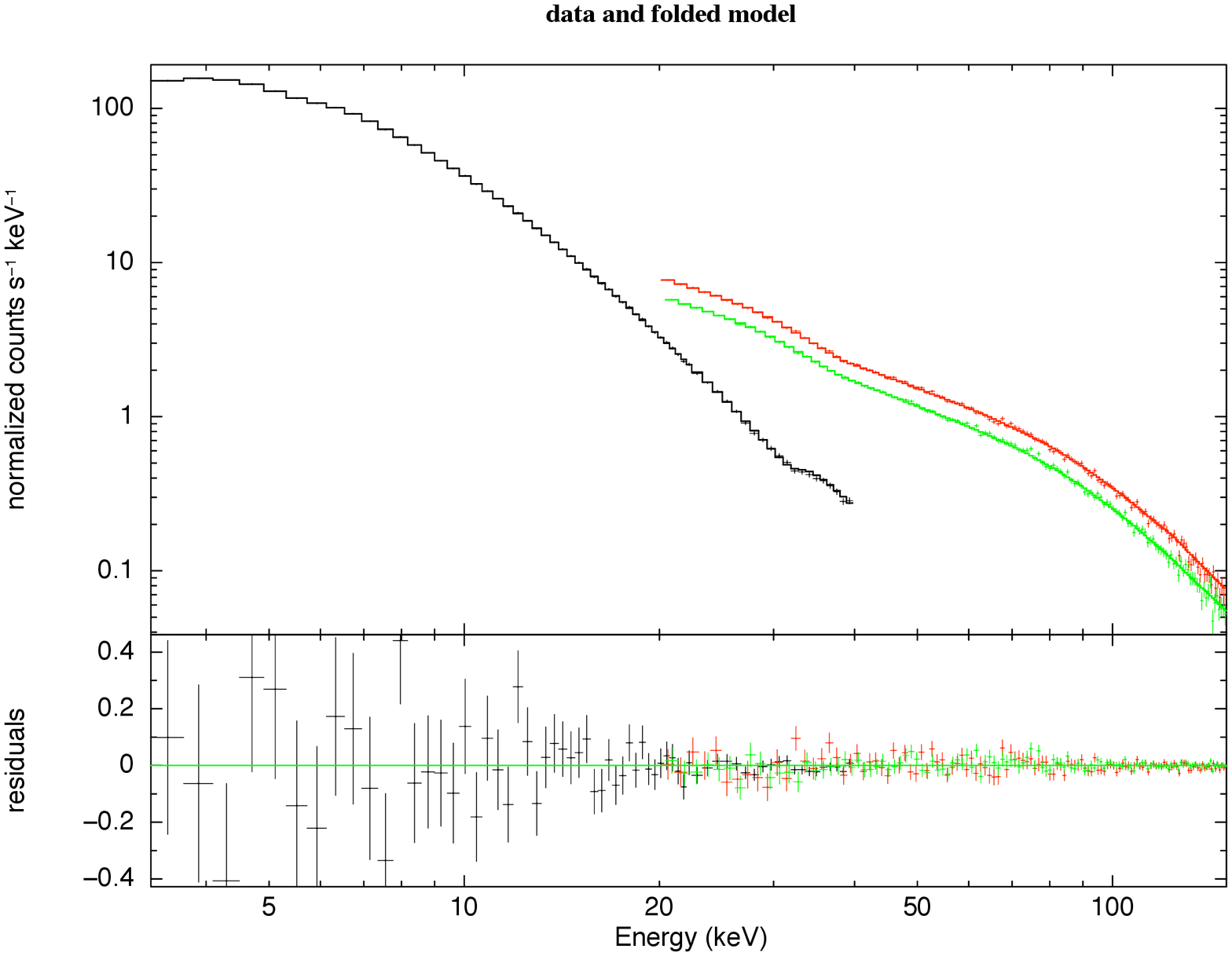}\\
Obs.~2\\
\includegraphics[height=6cm]{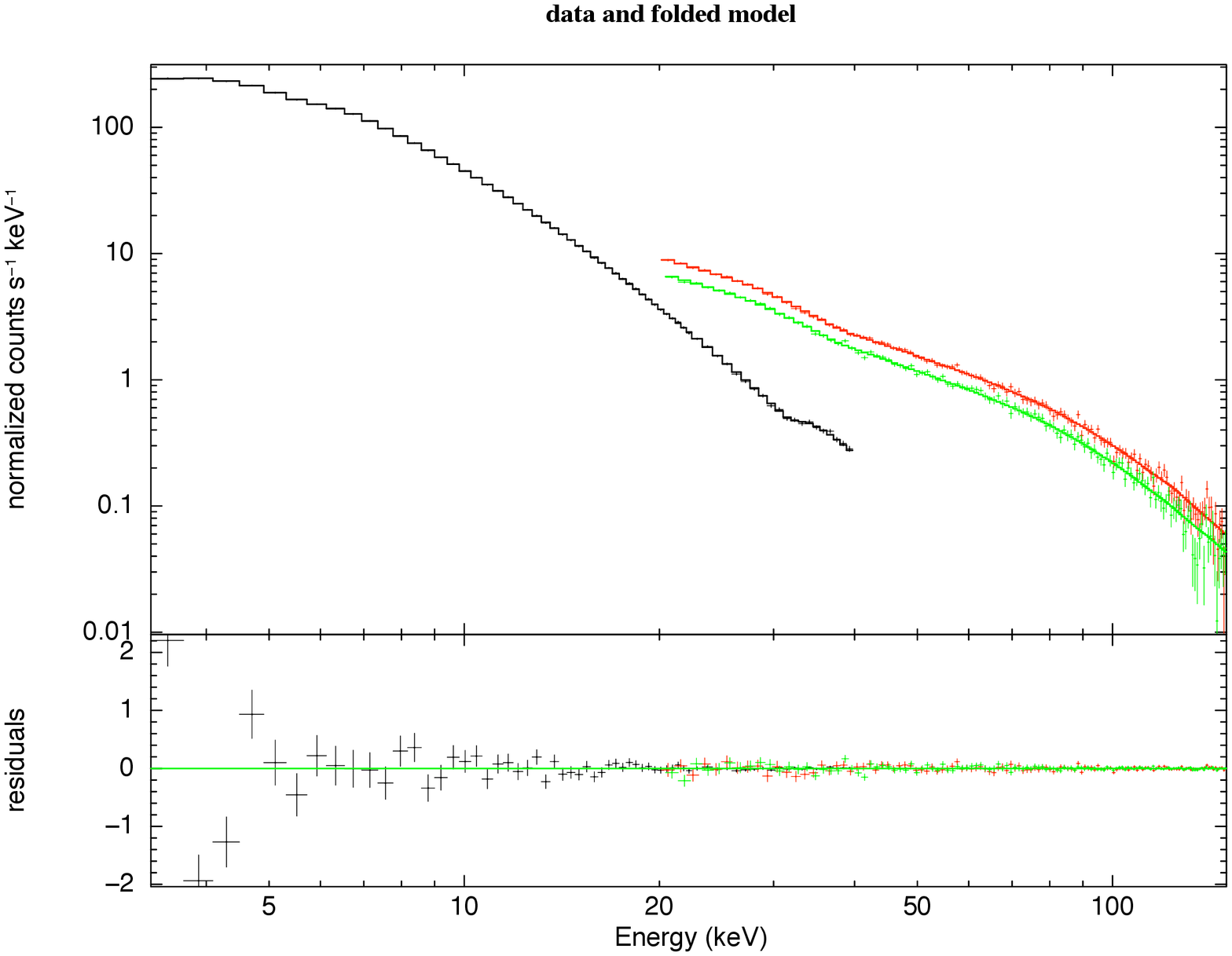}\\
Obs.~3\\
\includegraphics[height=6cm]{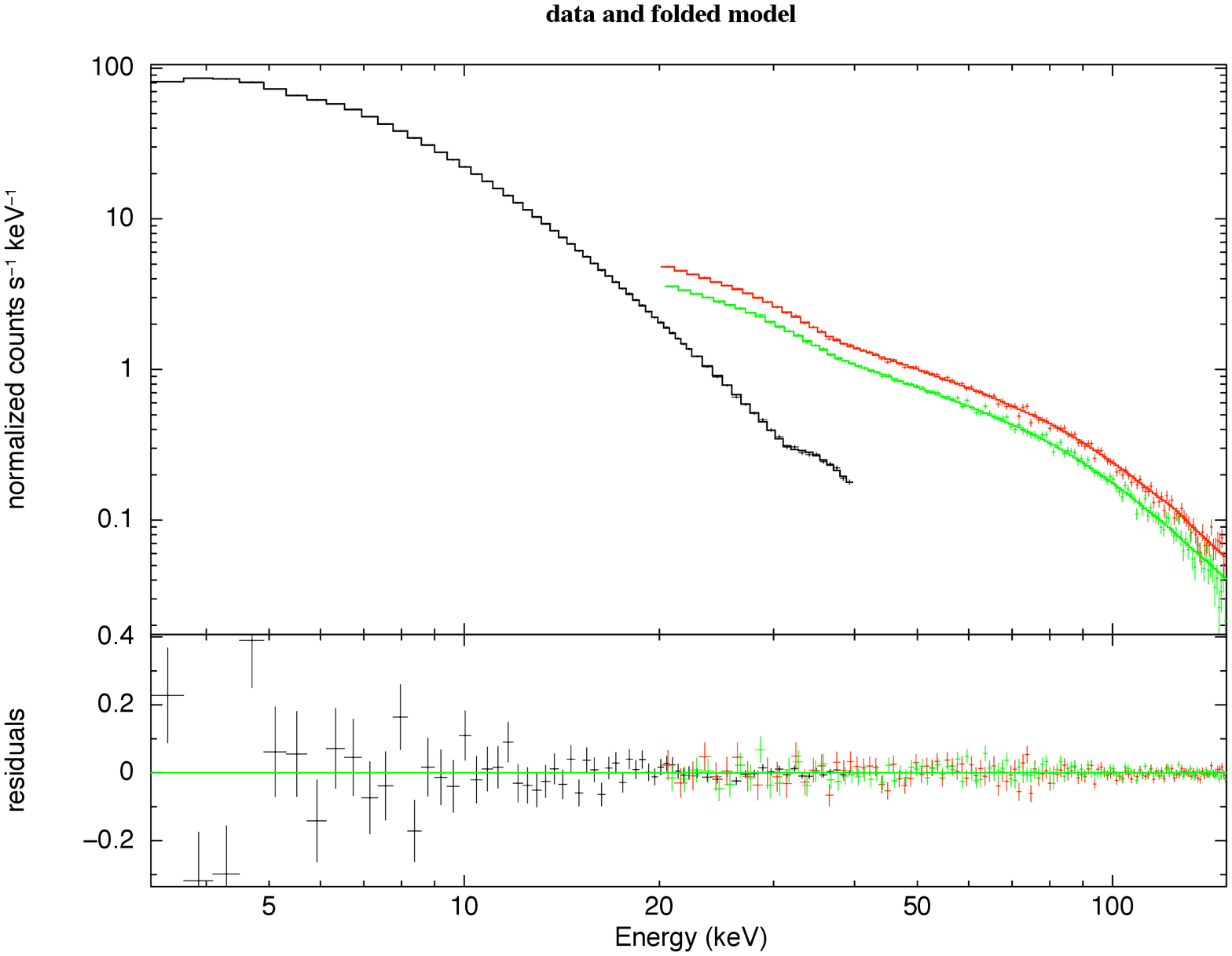}\\
\end{tabular}
\caption{\small Obs.~1, Obs.~2, and Obs.~3 \rxte/PCA+HEXTE spectra of \cygx1\ fitted with the model \textit{phabs(highecut$\times$bknpower+gaussian}).}
\label{xfit}
\end{center}
\end{figure*}

The \rxte\ spectra are fitted between 3 and 200~keV in  {\tt XSPEC} v.12.5.1. We tested several continuum models such as the standard $\alpha-$disk \citep{1984Mitsuda} plus a Comptonization component \citep{1994Titarchuk} modified by photoelectric absorption, but we find that all the spectra are best described by the phenomenological model combining a broken power law with a high energy cut-off and a gaussian to account for the iron emission, both modified by photoelectric absorption. These results are consistent with previous studies \citep[see e.g.][]{2006Wilms}. The best-fit parameters for all the observations are listed in Table~\ref{xpar} and the fits are displayed in Fig.~\ref{xfit}. 

Based on the classification of \cygx1's spectral states as given by \citet{2006Wilms}, these fit results show unambiguously that the source is in the HS during Obs.~1 and Obs.~3, with $\Gamma_1$ photon indices of about 1.83 and 1.75, respectively. The $3-200$~keV unabsorbed flux is nevertheless about 50\% larger during Obs.~1 than during Obs.~3 which, considering the X-ray/radio correlation observed in microquasars in the HS, suggests a stronger contribution of compact jets to the overall emission of \cygx1\ during Obs.~1. In contrast, the spectrum of the source during Obs.~2 is softer, with $\Gamma_1\approx2.08$. Such a photon index is consistent with an IS. But similarly, the high $3-200$~keV unabsorbed flux suggests a stronger contribution of compact jets than during Obs.~3.  

\subsection{Light curves}

\begin{figure}
\begin{center}
\includegraphics[width=15cm]{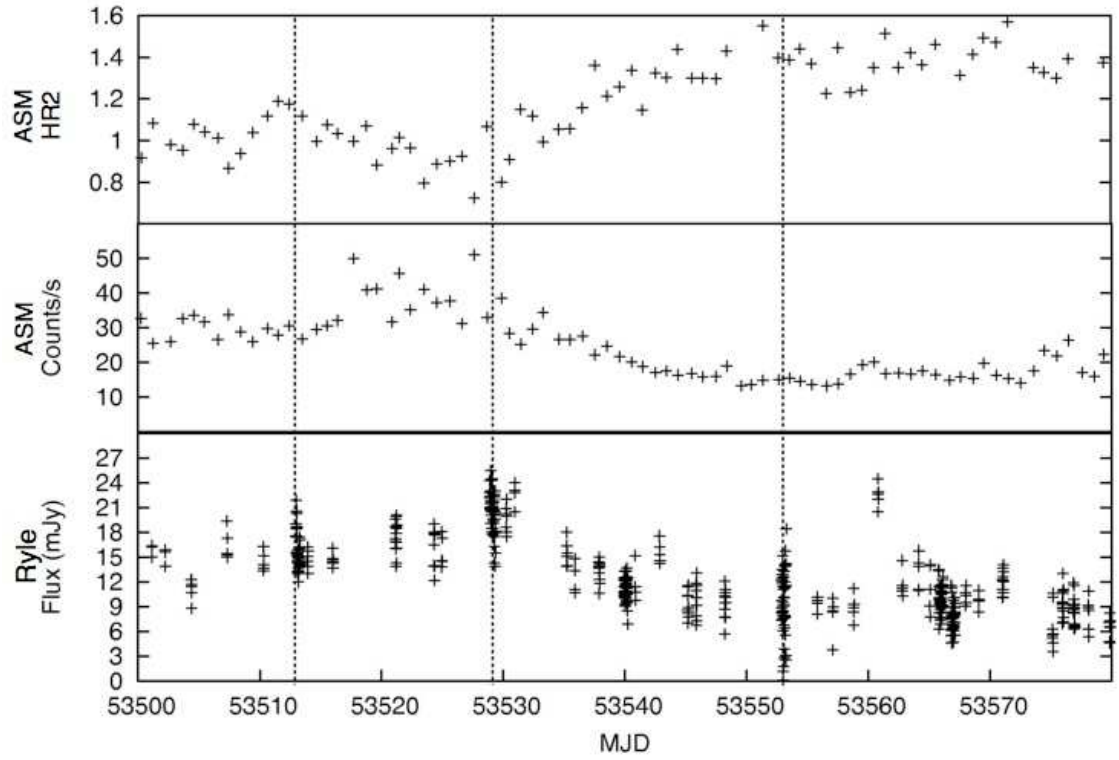}
\caption{\small Ryle (15~GHz, bottom) and  ASM (1.2$-$12.0~kev, middle) light curves, as well as the ASM HR2=C/A hardness ratio (top) of \cygx1\ between MJD~53500 and MJD~53580. Our three IRS spectroscopic observations are marked in dashed lines.}
\label{lc}
\end{center}
\end{figure}

\begin{figure}
\begin{center}
\includegraphics[width=15cm]{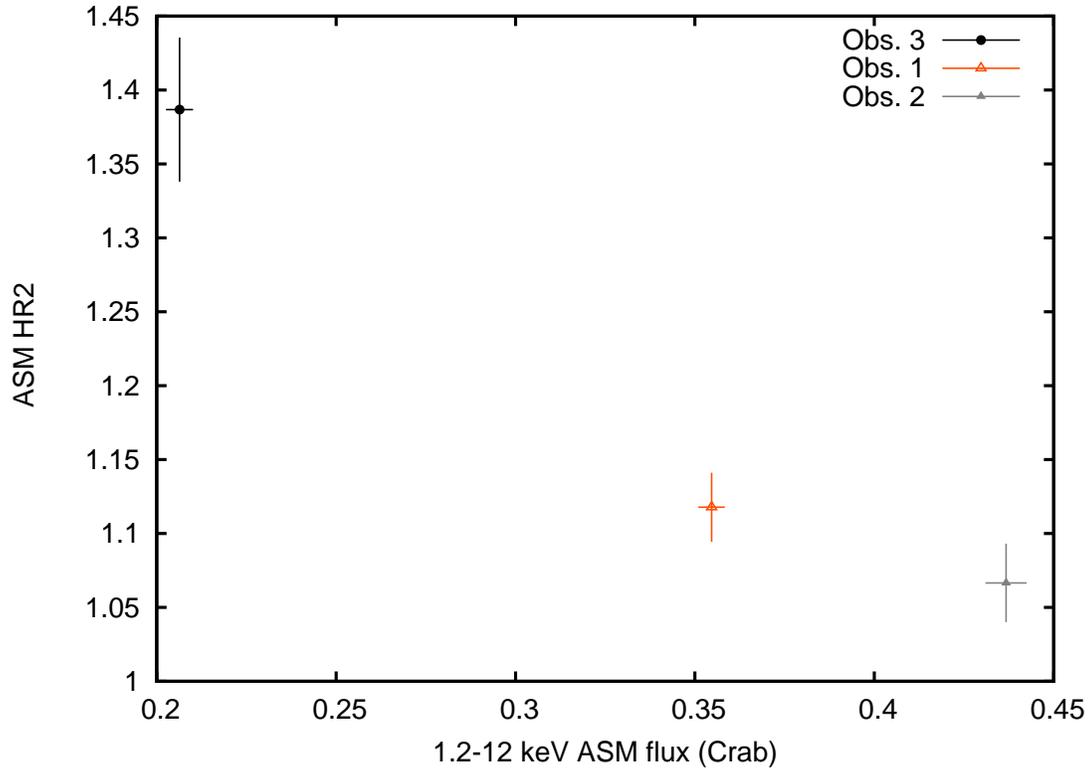}
\caption{\small Hardness-Intensity Diagram of \cygx1\ corresponding to our three observations. The X-axis is the 1.2$-$12.0~kev ASM flux in Crab ($1~{\rm Crab}=75.5$~counts/s), while the Y-axis is the ASM HR2=C/A hardness ratio. Clearly, Obs.~3 is the faintest and hardest observation, while Obs.~2 is the brightest and softest.}
\label{hid}
\end{center}
\end{figure}

In Fig.~\ref{lc}, we display the \cygx1\ radio (15~GHz, 5~min time resolution) and ASM (1.2$-$12~keV, day-averaged) light curves, as well as the ASM HR2 (C/A) hardness ratio, between MJD~53500 and MJD~53580, bracketing our observations. Clearly, it is seen that the count rate is lower and the hardness ratio higher during Obs.~3, i.e. when the MIR continuum appears to be at its lowest level (see Fig.~\ref{specdered}, blue spectrum). Consistent with the spectral fitting, the ASM light curve and HR2 hardness ratio show that the source during Obs.~2 is in a softer state and exhibits a slightly higher flux than during Obs.~1 (see also Fig.~\ref{hid}). This is in good agreement with the light curve trend shown in Fig.~\ref{lc} between MJD~53510 and MJD~53530, which shows a sharp decrease in the hardness ratio and a slow increase in the count rate. This probably points to a higher accretion disk activity becoming unsteady around MJD~53520. Then, the count rate drops between MJD~53530 and MJD~53550 before stabilizing, while the hardness ratio follows the opposite path. This behavior is characteristics of a failed-state transition as characterized by \citet{2006Wilms}. 

\begin{figure*}
\begin{center}
\begin{tabular}{ccc}
Obs.~1\\
\includegraphics[height=6cm]{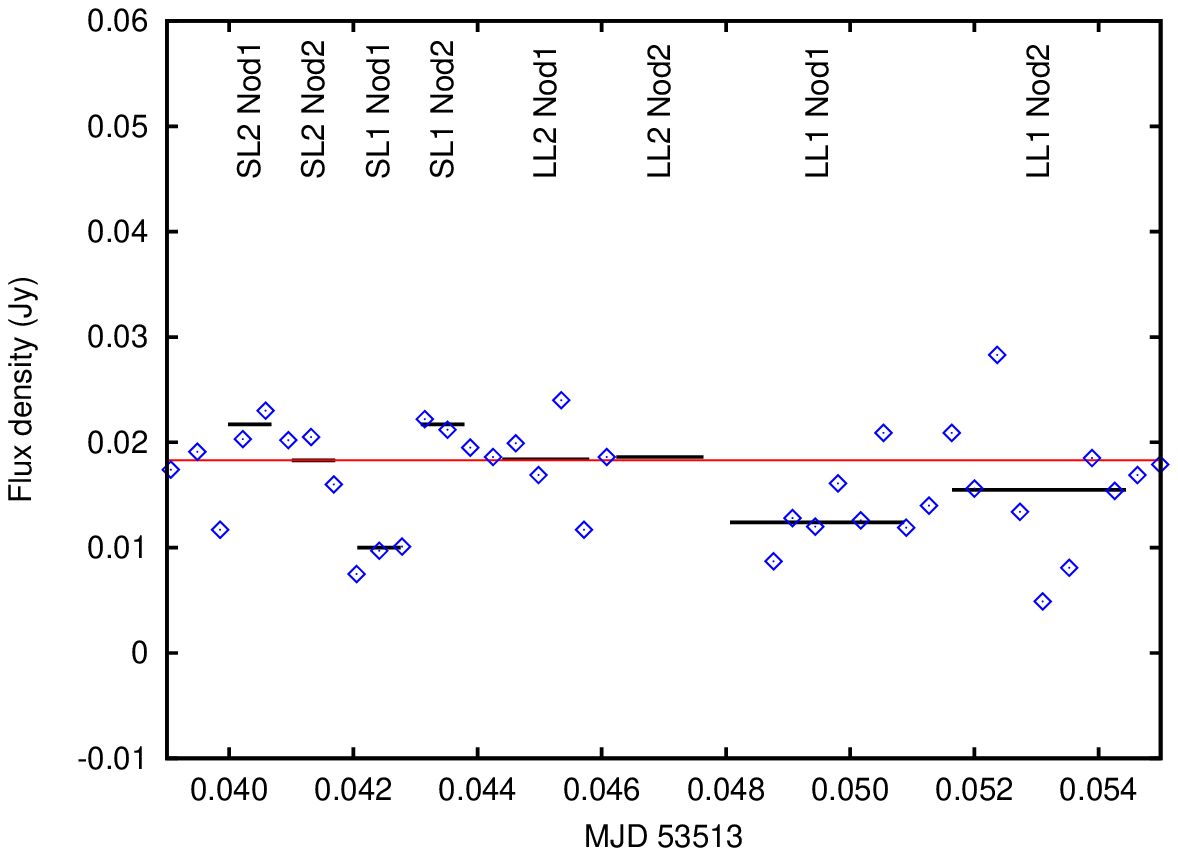}\\
Obs.~2\\
\includegraphics[height=6cm]{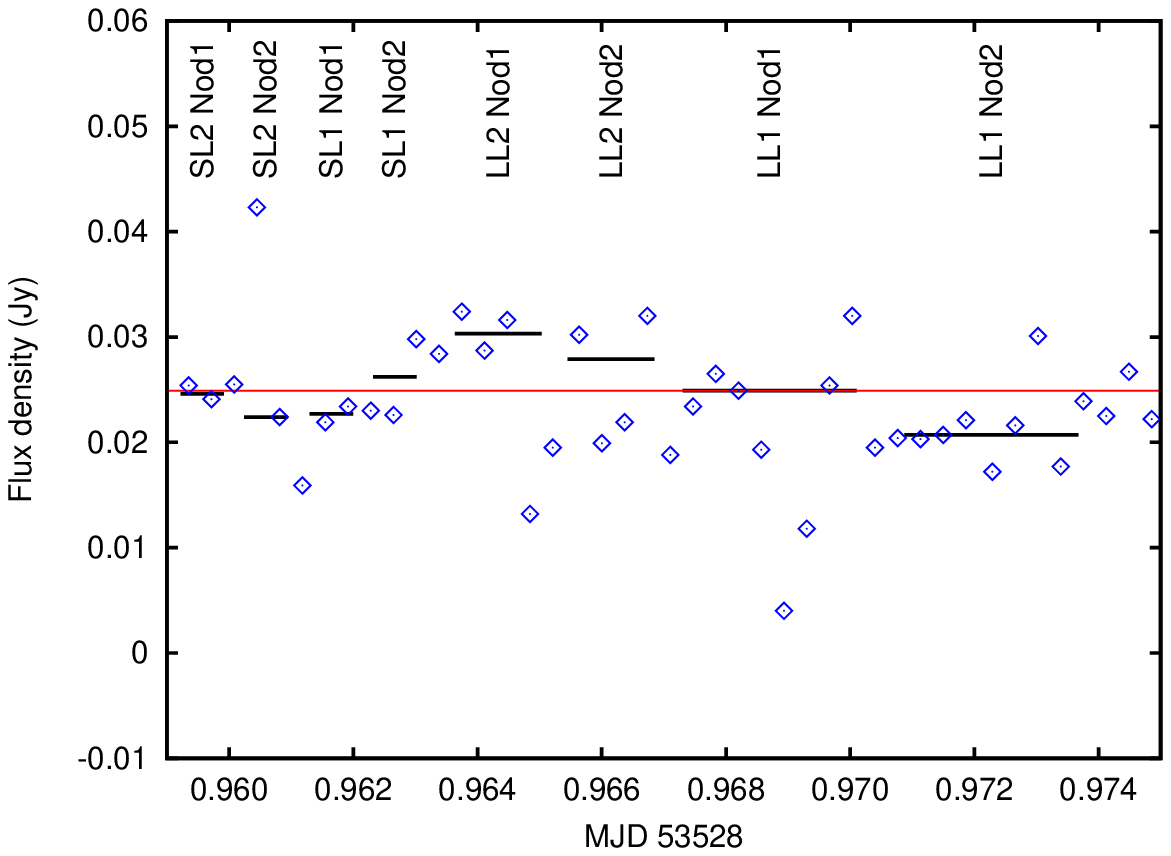}\\
Obs.~3\\
\includegraphics[height=6cm]{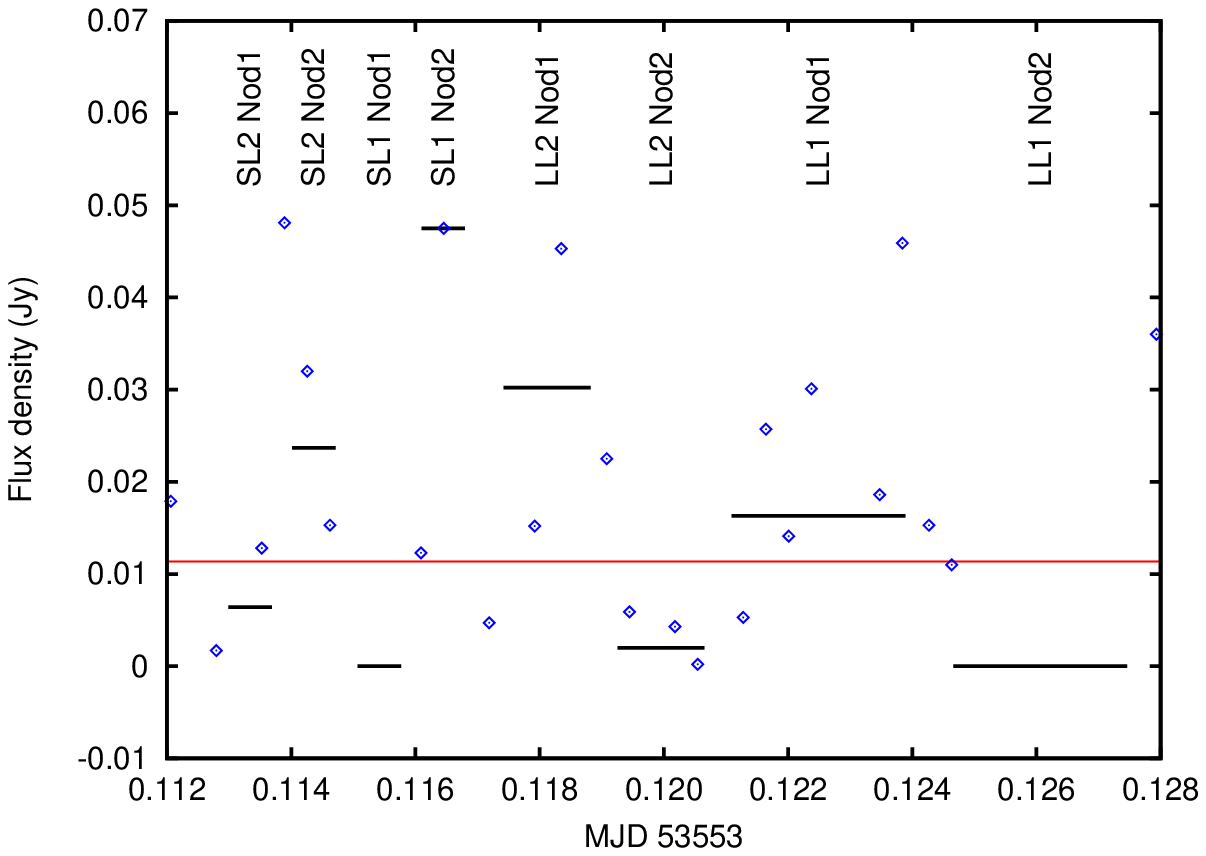}\\
\end{tabular}
\caption{\small Obs.~1, Obs.~2, and Obs.~3 Ryle light curves (20~s, blue diamonds) during each IRS integration. The black segments represent each module, the height being the median radio flux and the length the exposure time. The red line is the median radio flux during the whole observation.}
\label{radirs}
\end{center}
\end{figure*}

Moreover, the Ryle light curves during Obs~1. and Obs.~2 exhibit an average radio flux between 12 and 25~mJy, while the radio flux is on average lower and less steady during Obs.~3. This behavior is shown by the 20~s time resolution Ryle light curves, displayed in Fig.~\ref{radirs}, which cover the exact time range of each IRS observation. It is clear that Obs.~1 and Obs.~2 are consistent with compact jets. In contrast the radio activity during Obs.~3 is very unstable, covering a range from non-detection to 48~mJy. This behavior is not by itself inconsistent with the presence of compact jets, and possible explanations of this flickering could be attributed to some instabilities internal to the jet or some peculiar accretion flow properties. 

Finally, it is worth mentioning that there have been indications, from the timing behavior, of two different HS regimes \citep{2003Pottschmidt, 2008Axelsson}. One of them is characterized by a very high hardness level, a low X-ray flux, and characteristic frequencies shifted to the lower edge in the power density spectrum (PDS). This is clearly the case of the HS detected during Obs.~3 (see Grinberg et al. in prep, for the full PDS), which is actually the beginning of the hardest X-ray activity ever observed for \cygx1\ \citep{2011Nowak}. In contrast, the HS properties during Obs.~1 are consistent with what is expected in the canonical picture. 

\section{The MIR continuum fitting}

We use spectral fitting to understand the origin of the continuum variation between the three IRS observations. Between nod~1 and nod~2, we on average found 4\% and 5\% flux differences in SL and LL modules, respectively. These quantities were quadratically added as systematic errors to take into account uncertainties due to flux calibration. Since \cygx1\ was never in the SS during our observations, we rule out bremsstrahlung from the accretion disk wind, and we only consider thermal emission from the companion star's photosphere, bremsstrahlung from the expanding stellar wind, and/or synchrotron from the compact jets. Moreover a bump is present in the three spectra between about 17 and 22~\mic. The origin of this bump is a matter of debate, and it could be due to the dereddening process or to irradiation of dust around the BH (S. Markoff, private communication), but more likely to the \spitzer/IRS \textit{LL1 24~\mic\ deficit}\footnote{http://ssc.spitzer.caltech.edu/irs/features/\#3\_LL1\_24u\_Deficit}. 

\subsection{Obs.~3: the stellar continuum}

Considering the very unstable radio activity of \cygx1\ during Obs.~3, a jet contribution to the MIR spectrum is very unlikely. We therefore only consider the emission from the star. To model the photosphere, we simply use a black body. \cygx1's companion star, HD~226868, is an O9.7ab supergiant and its unabsorbed emission peaks in the ultraviolet (UV). The MIR alone is therefore not sufficient to constrain the star's temperature, so we fix it to the most recent derived value, i.e. 28000~K as given in \citet{2009Caballero}, based on theoretical fitting to the ultraviolet spectrum. In their seminal papers, \citet{1975Wright} and \citet{1975Panagia} showed that the bremsstrahlung from a homogenous spherically expanding wind would be responsible for a MIR and radio excess $\propto \nu^{0.6}$ in supergiant stars. We therefore add a power law to take the wind emission into account. This simple model is summarized by Eq.~\ref{eqmirobs3}: 

\begin{equation}
F(\nu)=\left ( \frac{R_\ast}{D_\ast} \right )^2 \times B(\nu, 28000)+N_{\rm ff}\left (\frac{\nu}{15\textrm{~GHz}} \right )^{\alpha_{\rm ff}}
\label{eqmirobs3}
\end{equation}
where $R_\ast$ and $D_\ast$ are the companion star's radius and distance, $B$ is the black body emission, and $N_{\rm ff}$ and $\alpha_{\rm ff}$ are the bremsstrahlung flux density at 15~GHz and spectral index.

\begin{figure*}
\begin{center}
\includegraphics[height=8cm]{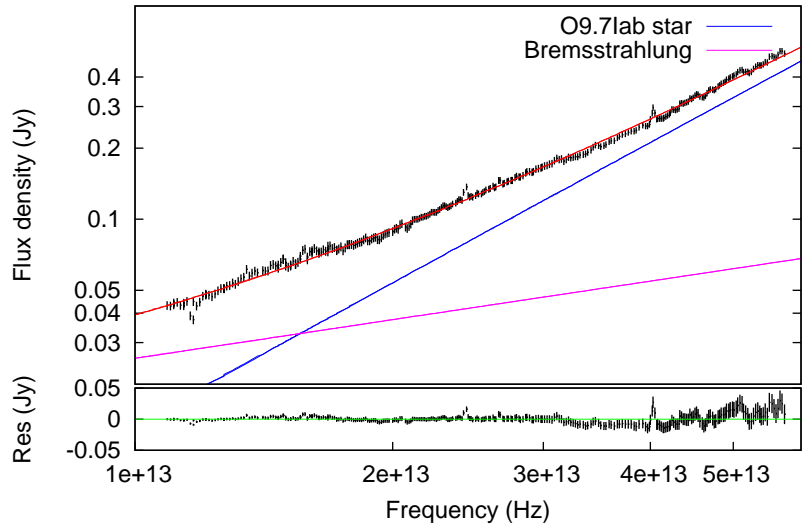}
\caption{\small Fit of the MIR continuum of \cygx1\ during Obs.~3, with a stellar black body and an expanding wind bremsstrahlung.}
\label{mirobs3}
\end{center}
\end{figure*}

The best fit is displayed in Fig.~\ref{mirobs3} and the best-fit parameters are listed in Table~\ref{mirparobs3}.  The ratio $R_\ast/D_\ast$ is consistent with the one expected from the \cygx1's companion star, since the derived stellar radius for a typical 2~kpc distance is about 19.9~\rsun, the typical value for a typical late O supergiant. Moreover, the power law spectral index is characteristics of bremsstrahlung from an expanding stellar wind, with $\alpha_b\approx0.55$, consistent with the 0.6 theoretical value within the 90\% uncertainties. The MIR continuum of \cygx1\ during Obs.~3 can therefore be described by that of HD~226868. This alone does not rule out the possible presence of compact jets. Nevertheless, both the average X-ray and radio fluxes are at least 50\% lower than during Obs.~1 and 2. So, even if they are present, either the compact jets are too faint to be detected in the MIR, or they break to the optically thin regime beyond 30~\mic. Indeed, in this regime, the IR flux is expected to vary with respect to the radio flux as $F_{IR} \approx {F_{radio}} ^{21/17}$ \citep{2004Heinz}. Considering the low radio flux level, this very probably excludes a significant contribution of a compact jet to the MIR spectrum of \cygx1\ during Obs.~3.

\subsection{Obs.~1 and Obs.~2: MIR contribution from the compact jets}

The radio behavior does exhibit compact jets that may be responsible for the observed continuum increase. We consider two cases to model their contribution, (1) the spectral break does not occur in the IRS spectral range so the optically thick synchrotron emission can be described by a simple power law, or (2) the spectral break is located in the IRS spectral range and the emission of the compact jets can be modeled by a broken power law. 

To better constrain the compact jets emission parameters, we include the average 15~GHz flux densities of \cygx1\ during the IRS/LL observations. We then fit these MIR/radio spectral energy distributions (SEDs) by fixing the black body parameters and bremsstrahlung spectral index to those of Obs.~3.

\subsubsection{Case 1: simple power law}

The overall equation describing the \cygx1\ emission is: 

\begin{equation}
F(\nu)=\left ( \frac{R_\ast}{D_\ast} \right )^2 \times B(\nu, 28000)+N_{\rm ff}\left (\frac{\nu}{15\textrm{~GHz}} \right )^{\alpha_{\rm ff}}+N_{\rm j}\left (\frac{\nu}{15\textrm{~GHz}} \right )^{\alpha_{\rm j}}
\label{eqmircase1}
\end{equation}
where $N_{\rm j}$ and $\alpha_{\rm j}$ are the synchrotron flux density at 15~GHz and spectral index, respectively.

\begin{figure*}
\begin{center}
\begin{tabular}{c}
Obs.~1\\
\includegraphics[height=8cm]{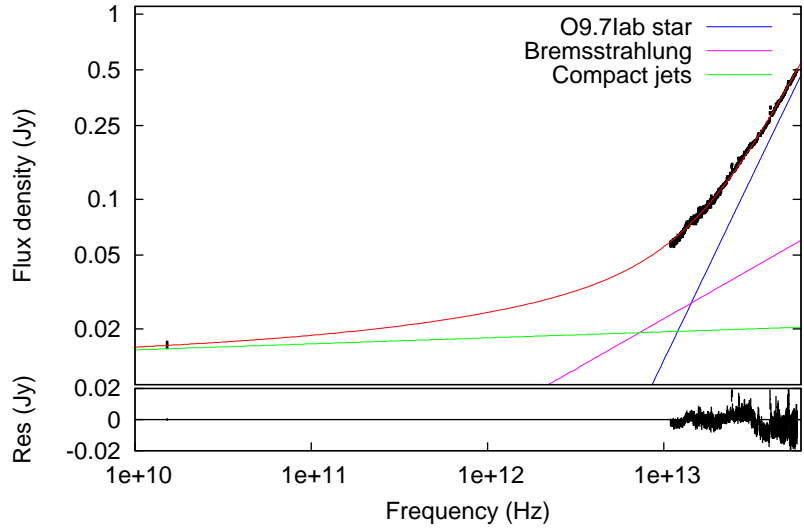}\\
Obs.~2\\
\includegraphics[height=8cm]{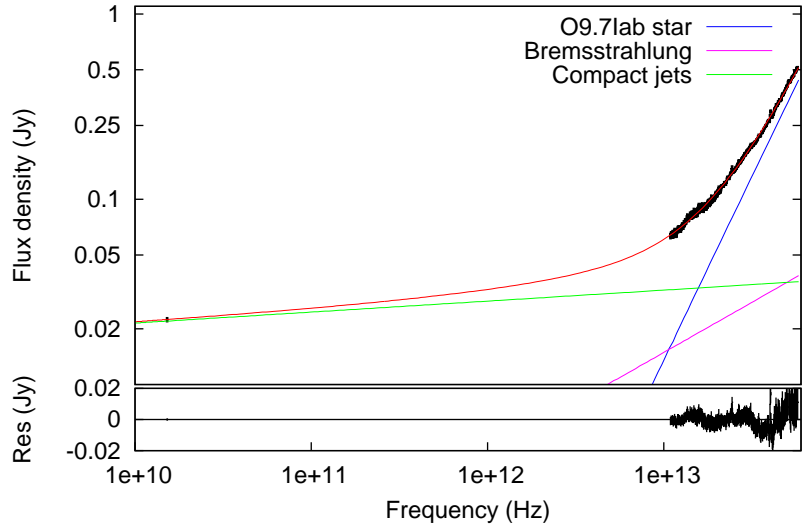}\\
\end{tabular}
\caption{\small Obs.~1 and Obs.~2 MIR/radio SEDs of \cygx1: case 1.}
\label{sedcase1}
\end{center}
\end{figure*}

Here, $N_{\rm ff}$ was left free to mimic a possible change in the bremsstrahlung intensity due to a variation of the star's mass loss rate. The best-fit parameters are listed in Table~\ref{mirparcase1}, and the corresponding fitted SEDs displayed in Fig.~\ref{sedcase1}. Both fits are consistent with a contribution of compact jets to the MIR emission of \cygx1\ during Obs.~1 and Obs.~2. Indeed, the derived spectral indices are $0.01\pm0.02$ and $0.05\pm0.01$, respectively, consistent with typical values. Moreover, the bremsstrahlung flux density at 15~GHz during Obs.~1 is found to be similar to that during Obs.~3 (the no-jet case, see Sect. 4.1), $0.66\pm0.06$~mJy compared to $0.74^{+0.10}_{-0.07}$~mJy. In contrast, $N_{\rm ff}$ during Obs.~2 is lower ($0.44^{+0.06}_{-0.05}$~mJy). 

\subsubsection{Case 2: broken power law}

Here, the overall equation describing the \cygx1\ emission is: 

\begin{equation}
F(\nu)=\left ( \frac{R_\ast}{D_\ast} \right )^2 \times B(\nu, 28000)+N_{\rm ff}\left (\frac{\nu}{15\textrm{~GHz}} \right )^{\alpha_{\rm ff}}+\left\{
    \begin{array}{ll}
        {\tiny N_{\rm j}}\left (\frac{\nu}{15\textrm{~\tiny GHz}} \right )^{\alpha_{\rm j1}}&\mbox{if } \nu \le \nu_{\rm c} \\
        {\tiny N_{\rm j}}\left (\frac{\nu_{\rm c}}{15\textrm{~\tiny GHz}} \right )^{\alpha_{\rm j1}-\alpha_{\rm j2}}\left (\frac{\nu}{15\textrm{~\tiny GHz}} \right )^{\alpha_{\rm j2}}&\mbox{if } \nu \ge \nu_{\rm c} \\
    \end{array}
\right.
\label{eqmircase2}
\end{equation}
where $N_{\rm j}$, $\alpha_{\rm j1}$, $\alpha_{\rm j2}$, and $\nu_{\rm c}$ are the jets' synchrotron flux density at 15~GHz, optically thick and thin spectral indices, and cut-off frequency, respectively.

\begin{figure*}
\begin{center}
\begin{tabular}{cc}
Obs.~1&Obs.~2\\
\includegraphics[width=8cm]{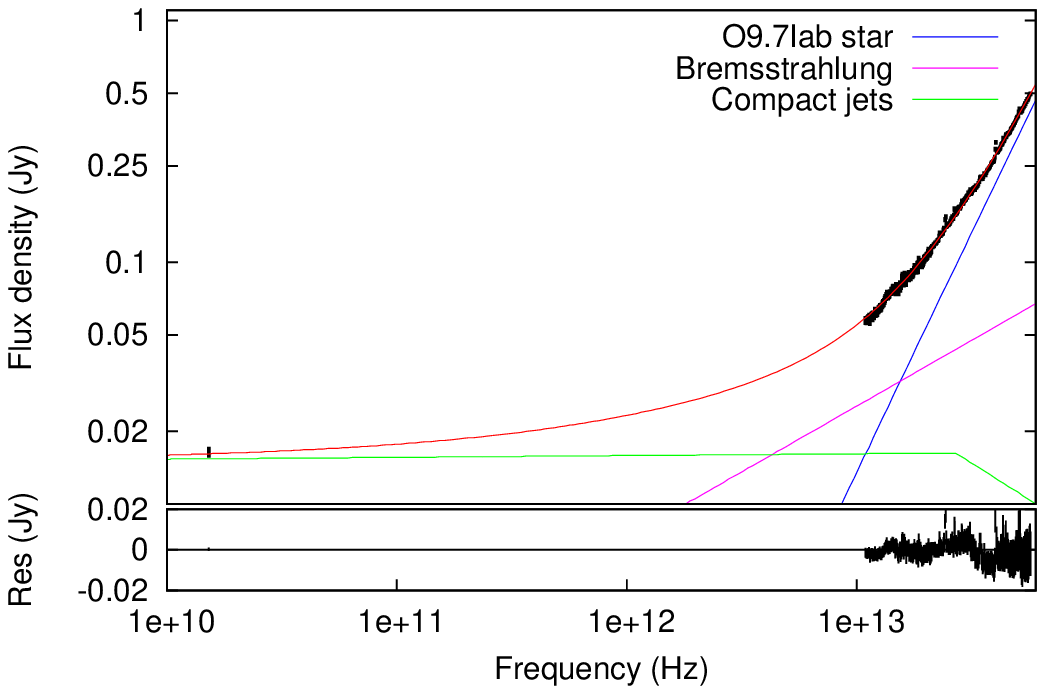}&\includegraphics[width=8cm]{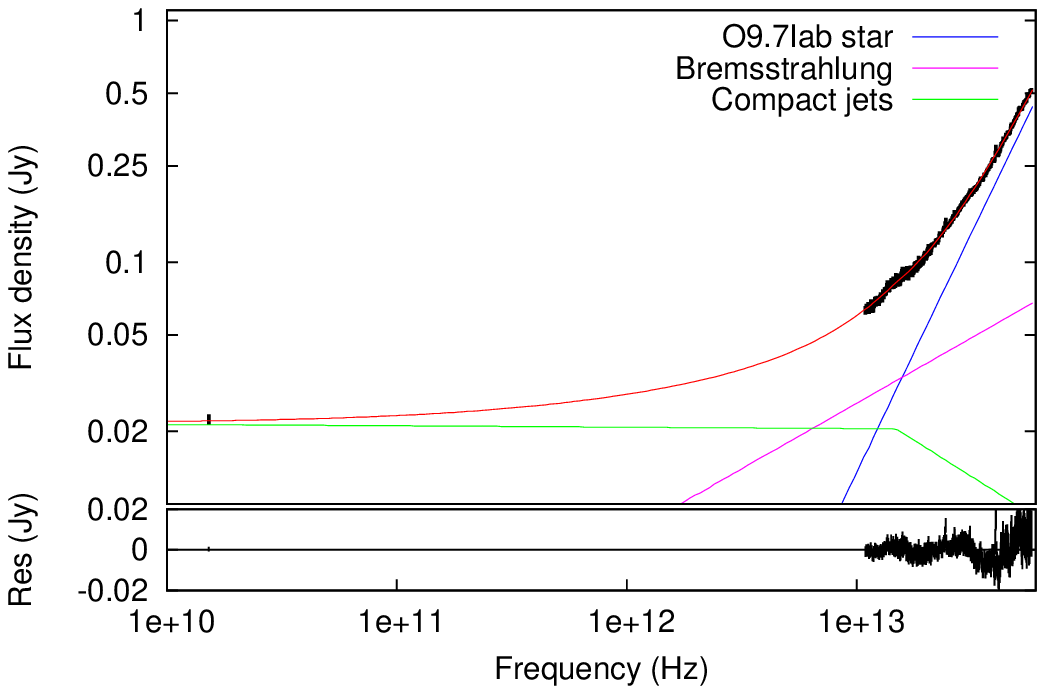}\\
\includegraphics[width=8cm]{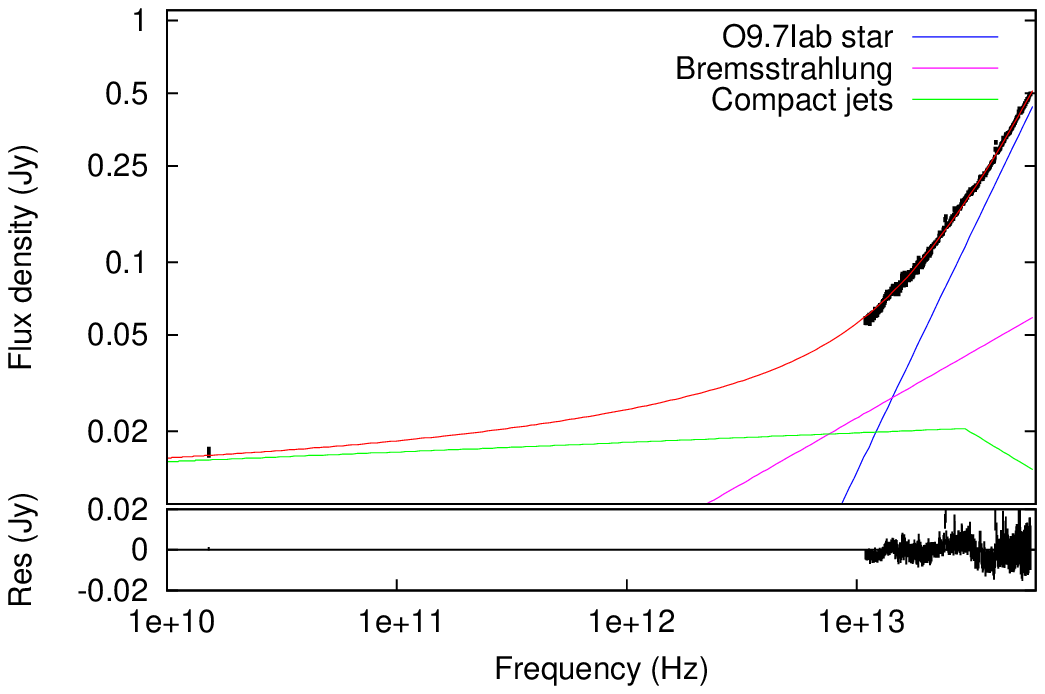}&\includegraphics[width=8cm]{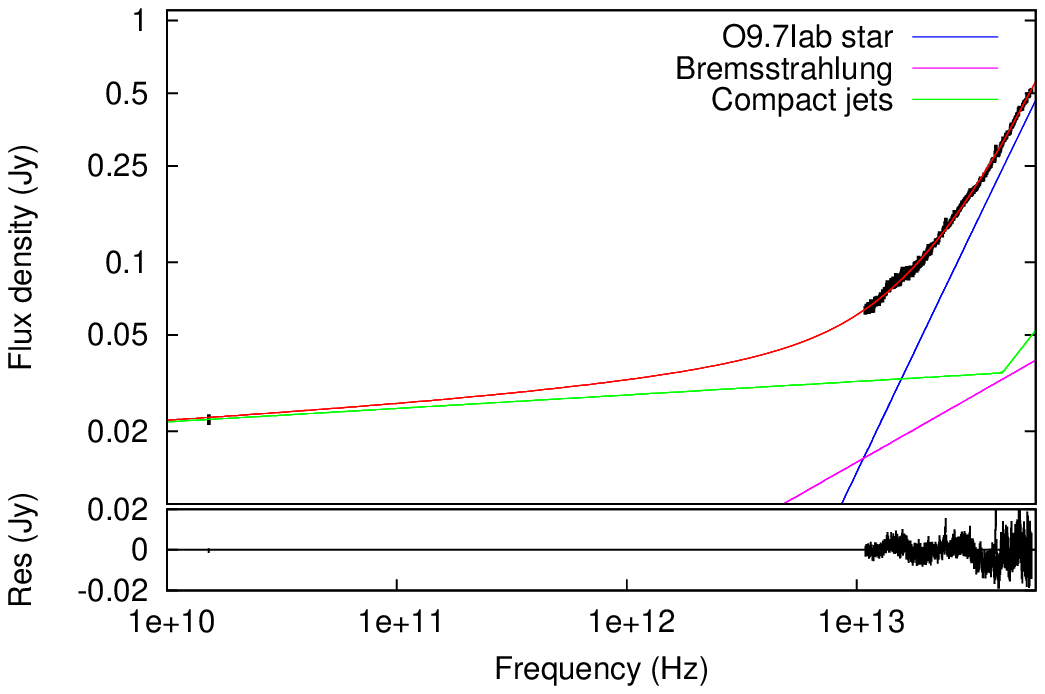}\\
\includegraphics[width=8cm]{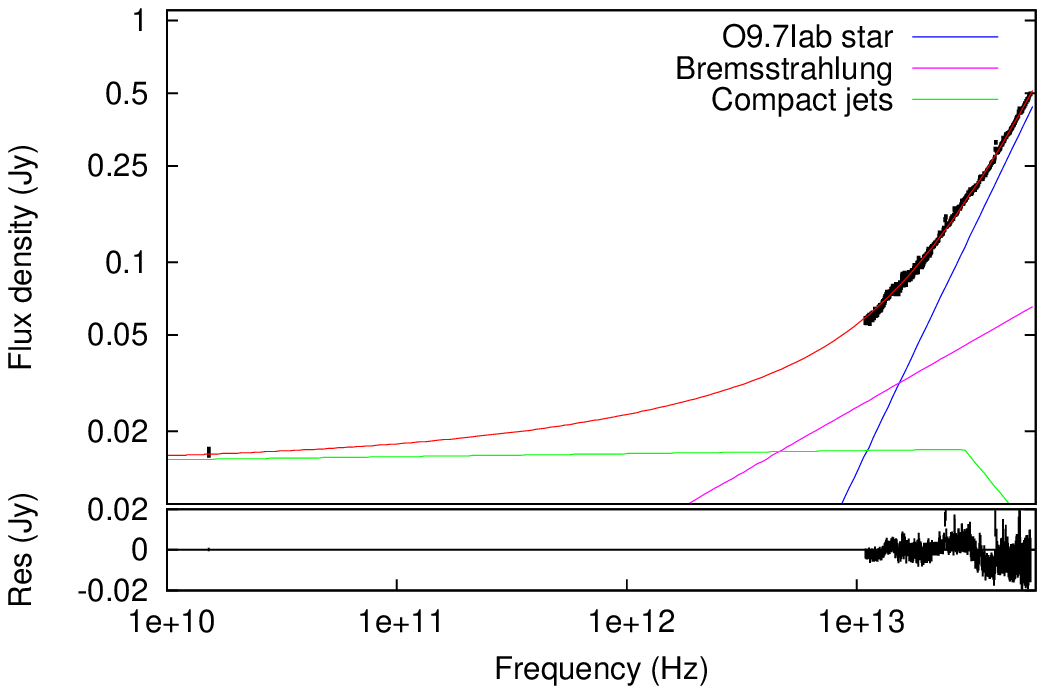}&\includegraphics[width=8cm]{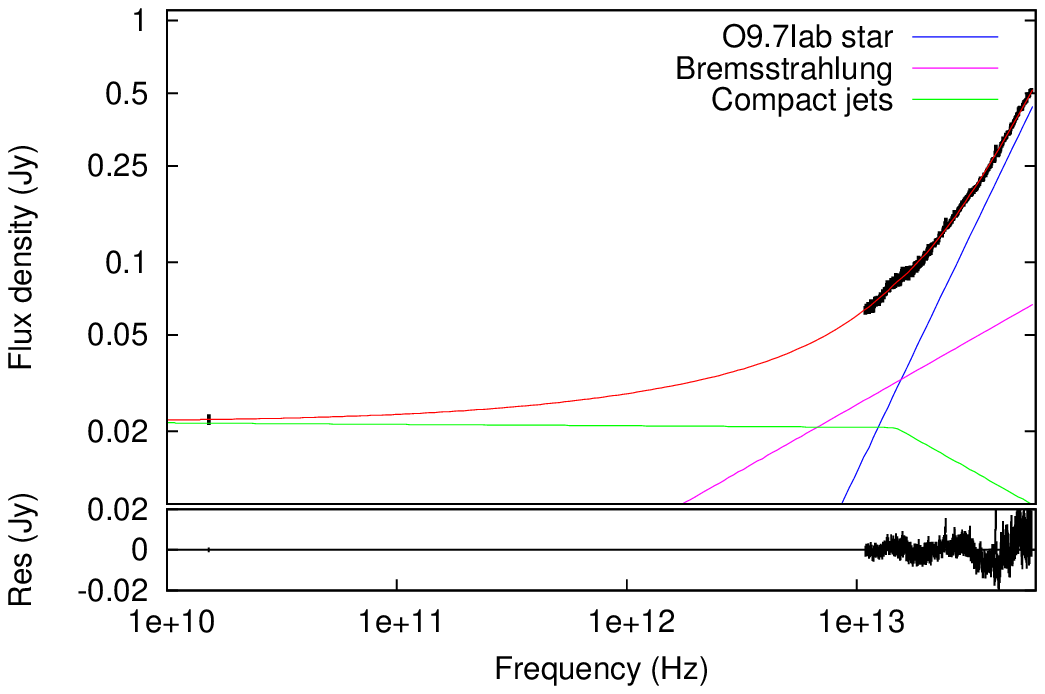}\\
\end{tabular}
\caption{\small Obs.~1 and Obs.~2 MIR/radio SEDs of \cygx1, case 2: (top) scenarios (a), (middle) scenario (b), (bottom) scenario (c).}
\label{sed1}
\end{center}
\end{figure*}

When we left free to vary all the parameters of the broken power law and the bremsstrahlung normalization $N_{\rm ff}$ at the same time; however this results in poorly constrained fits due to several degeneracies between the various fit parameters. We therefore proceeded with three scenarios, each time fixing one parameter only: (a) $\alpha_{\rm j2}$ is fixed to $-0.6$, i.e. the theoretical spectral index expected from the optically thin synchrotron emission of the compact jets, (b) $N_{\rm ff}$ is fixed to the values found in case~1, (c) $N_{\rm ff}$ is fixed to the value found during Obs.~3, i.e. we assume that the bremsstrahlung intensity is the same for all the observations. The best-fit parameters are listed in Tables~\ref{mirparcase21}, \ref{mirparcase22}, and \ref{mirparcase23}, and the SEDs are displayed in Fig.~\ref{sed1}, respectively. 

\subsubsection{Interpretation}

A broken power law gives better results to describe the emission of the compact jets during Obs.~1, with reduced \chis2\ values of 0.81, 0.85, and 0.78 for scenarios (a), (b), and (c), respectively, versus 0.96 for a simple power law. The low reduced \chis2\ are largely due to artificially forced high IRS systematic errors that were quadratically added to the uncertainties. Nevertheless, $\Delta\chi^2 \ge 34$ for only one DOF less, and the three fits give very similar results. The spectral break is clustered in the MIR range $(2.70-2.94)\times10^{13}$~Hz, i.e. about $10.06-11.11$~\mic, and the optically thin spectral index $\alpha_{\rm j2}$ for (b) and (c) is always consistent with the expected value (between $-1$ and $-0.4$) within the uncertainties. Moreover, for (a), the bremsstrahlung intensity is found similar to that during Obs.~3, a hint that the mass-loss rate is the same during both observations, as expected in the HS. So, based on the aforementioned reasons, we conclude that the consistency of all the fits strengthens the need for a broken power law to describe the MIR continuum of \cygx1\ during Obs.~1.

The interpretation of the results for Obs.~2 is more complex, even though the broken power law gives lower reduced \chis2\ (0.69 against 0.72).  First,  $\Delta\chi^2 \approx 9.7$, which is far less significant than the values found for Obs.~1. Moreover, the fits with the broken power law are not physically relevant. Indeed, the fit for scenario (b) gives a positive photon index after the spectral break ($\alpha\approx1.21$), which is clearly inconsistent with a compact jet. The two other fits are consistent with compact jets, but give cut-off frequencies that are about twice smaller than during Obs.~1. This is counter-intuitive since the opposite is expected. Indeed, the source during Obs.~2 is not in the HS but in a failed-state transition, i.e. an IS with a softer and brighter spectrum. In models of optically thick jet emission, the break frequency is coupled to the jet luminosity: if the luminosity increases, the break frequency also increases. An increase in the optically thick flux requires an increase in the jet energy density and/or the cross section of the jet, which always increases the optical depth, thus increasing the break frequency. The fact that our best fits for Obs.~2 indicate the opposite would imply a change in the jet geometry, e.g., a significant increase in the distance of the base of the jet along the jet axis or an increase in the opening angle of the jet (such that the footprint of the base of the jet shrinks, decreasing its optical depth). Given the level of uncertainty in our data (demonstrated by the fact that our models are under-constrained), such conclusions are still rather speculative at this point. Moreover, several studies of \cygx1\ showed the existence of an anti-correlation between the mass-loss and the accretion rates \citep{2003Gies, 2010Boroson}, maybe due to photoionization of the wind by the soft X-ray emission. During the IS and SS, the bremsstrahlung contribution should therefore be lower than in the HS. This is actually exactly the result of case~1, where we consider a simple power law instead of a broken one. The bremsstrahlung intensity is found to be about 40\% lower than that during Obs.~3, while the cut-off frequency is assumed to occur below 5.48~\mic. We believe that it better describes what occurs during Obs.~2.

\section{MIR spectral lines analysis}

\begin{figure*}
\begin{center}
\begin{tabular}{cc}
\includegraphics[width=8cm]{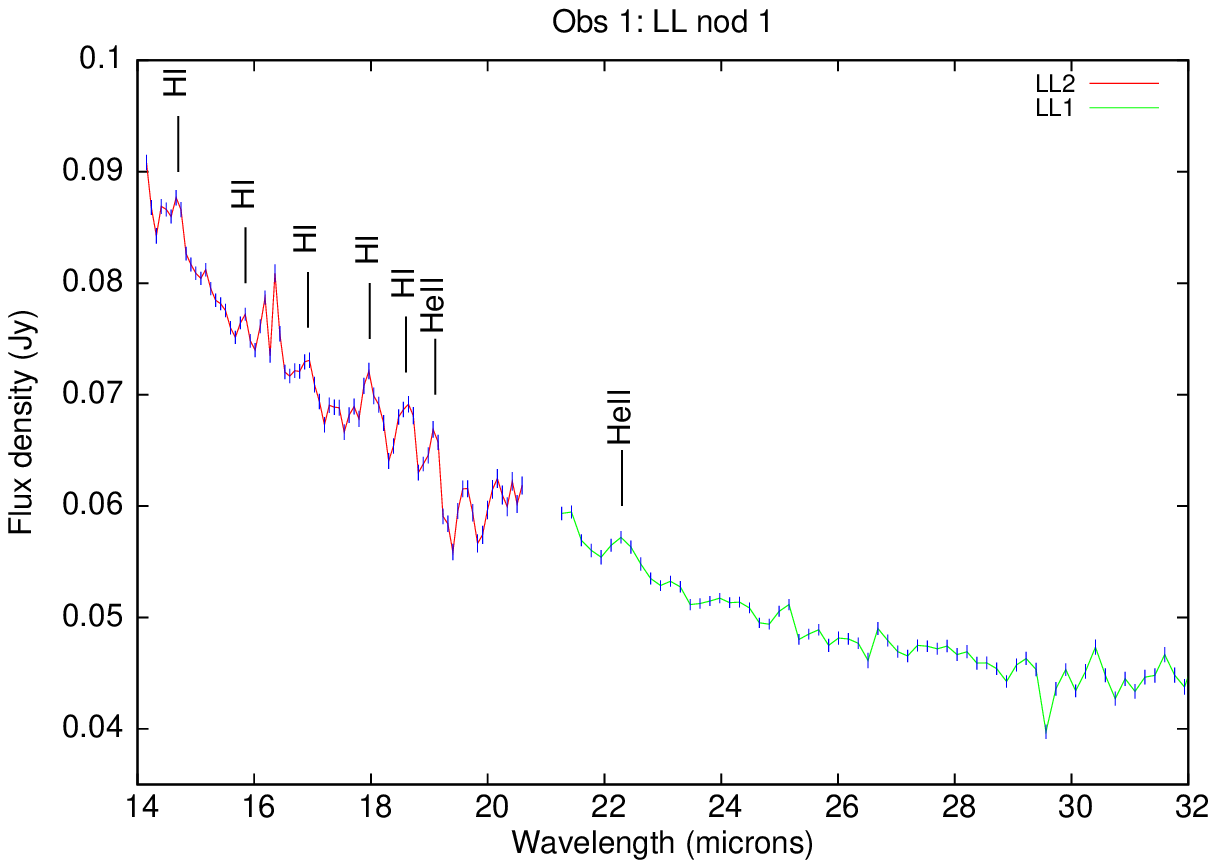}&\includegraphics[width=8cm]{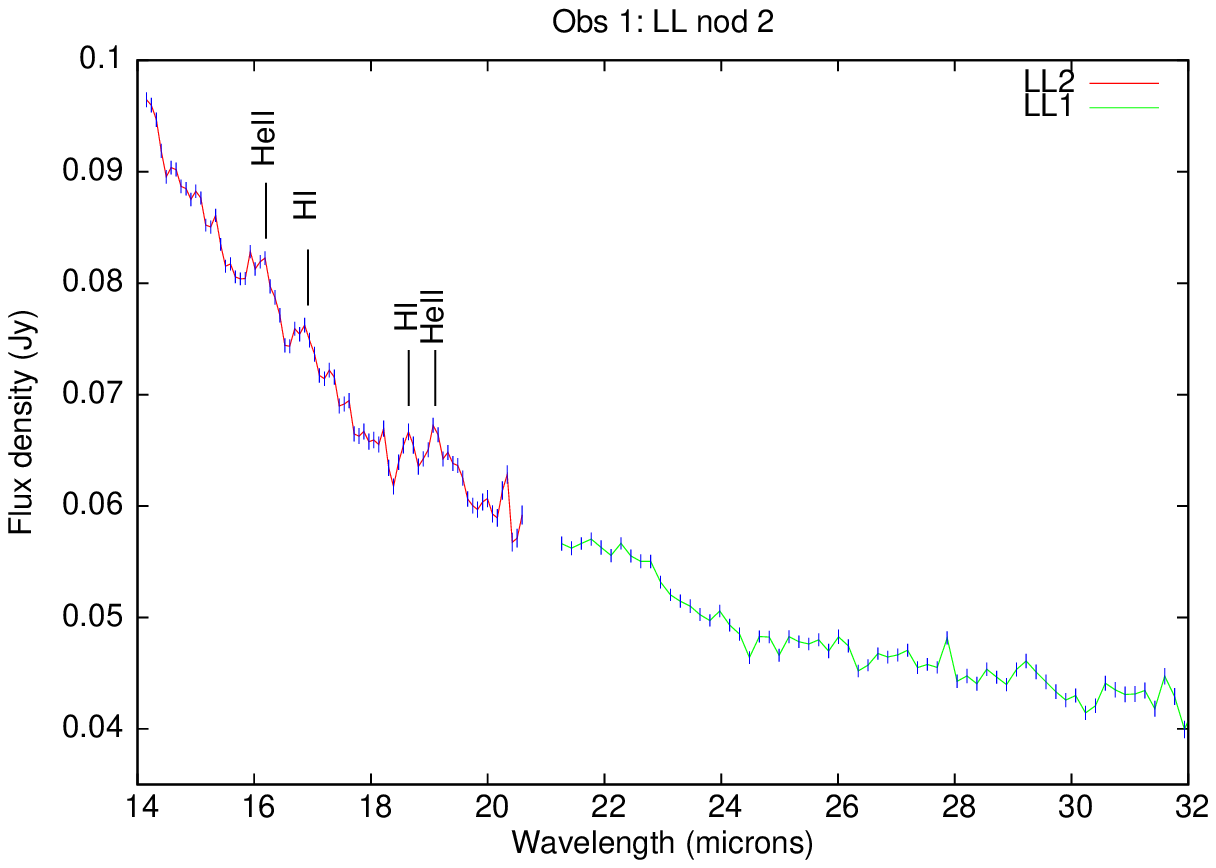}\\
\includegraphics[width=8cm]{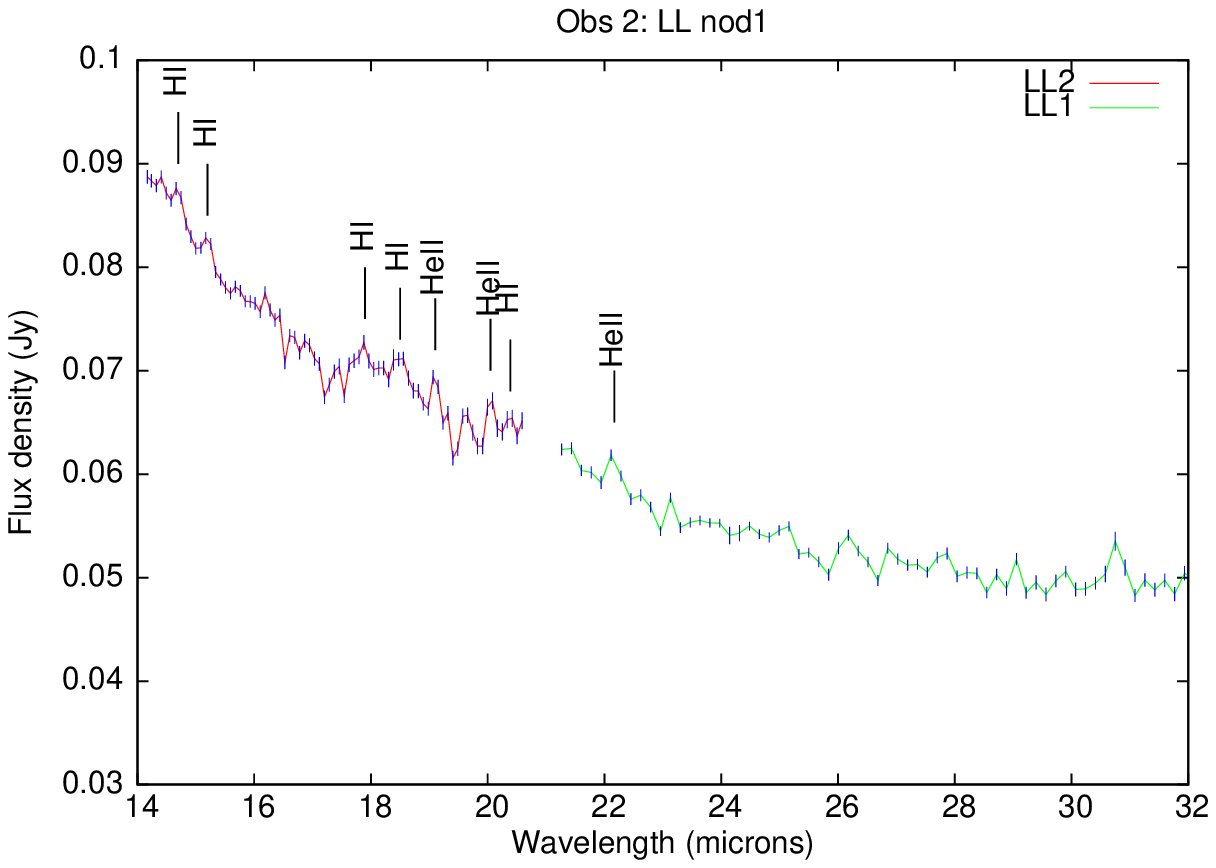}&\includegraphics[width=8cm]{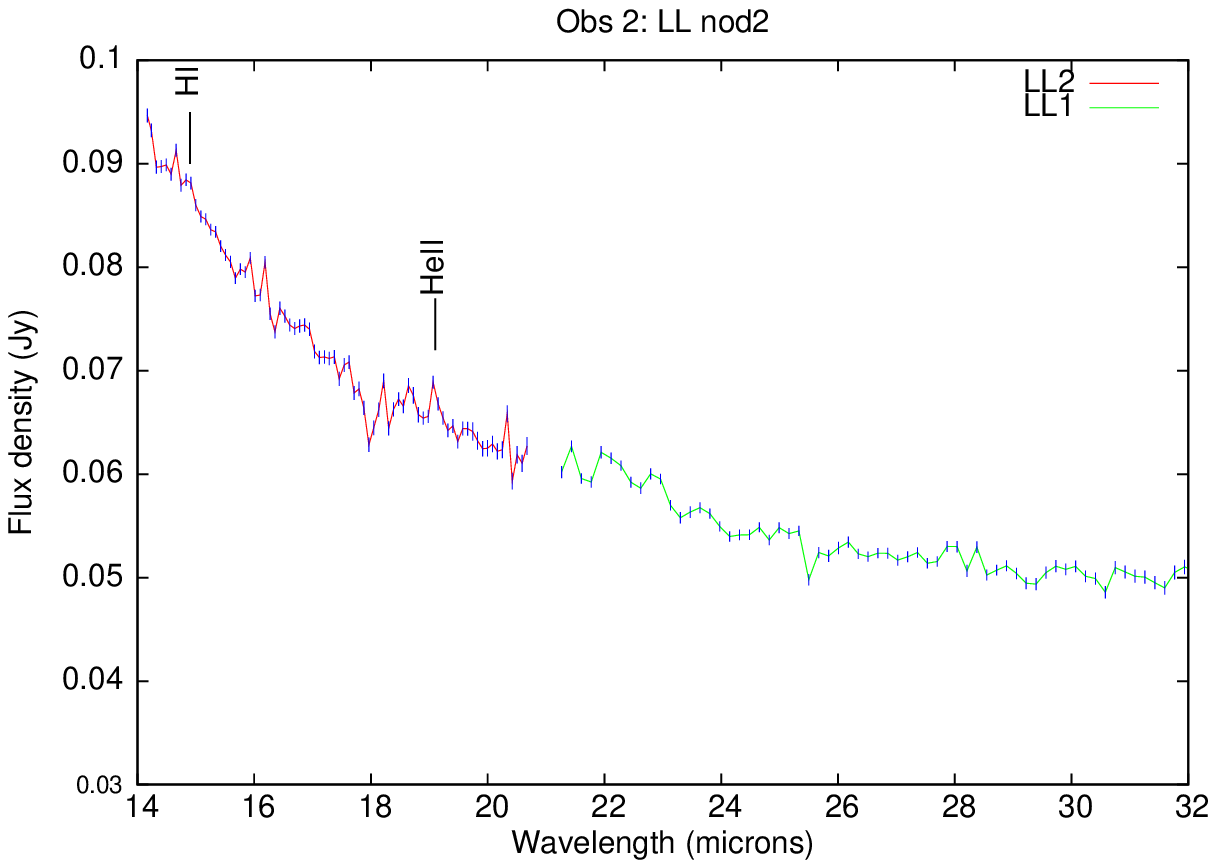}\\
\includegraphics[width=8cm]{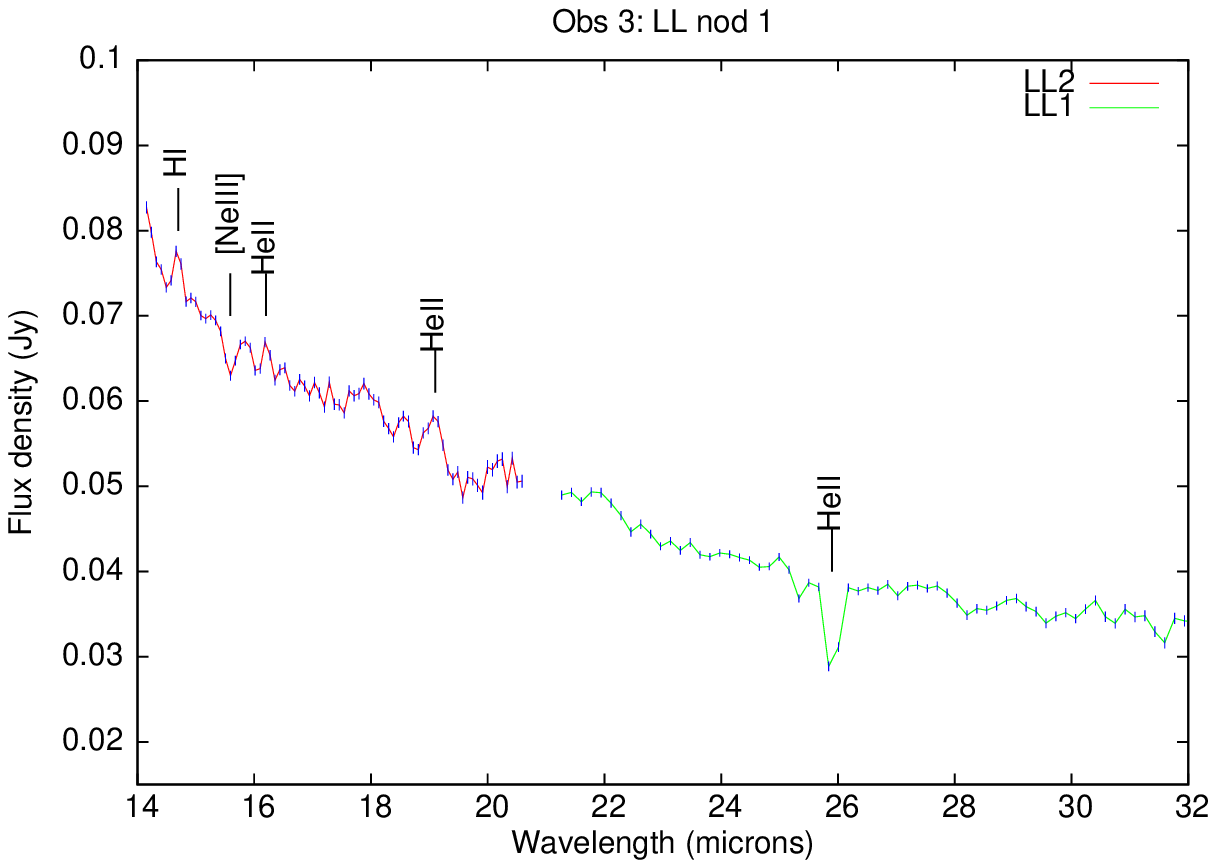}&\includegraphics[width=8cm]{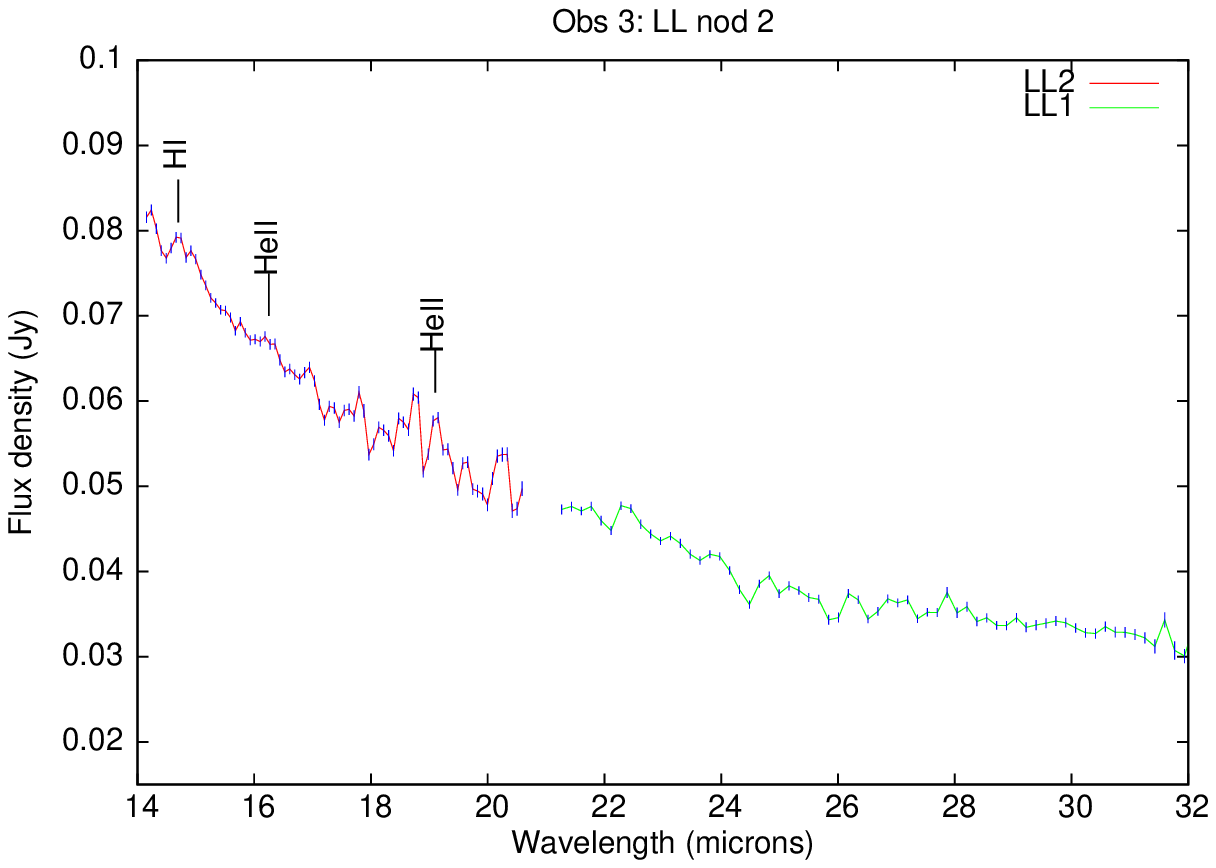}\\
\end{tabular}
\caption{\small LL nod 1 and nod 2 spectra of \cygx1\ during Obs.~1 (top), Obs.~2 (middle), and Obs.~3 (bottom).}
\label{spll}
\end{center}
\end{figure*}

In the companion paper (Lee et al., in prep), we show that the presence of warm dust around \cygx1\ is very unlikely and that the strong silicate absorption feature is due to the ISM. Here, we study the spectral line variability of the MIR spectra between the three different epochs. Moreover, because the spectral features are remarkably similar for the three spectra in the SL modules, we only focus on the longer wavelengths, from 14.5 to 32~\mic. Finally, due to potential rapid variations with respect to the source's X-ray or radio activity, we separately study nod~1 and nod~2, in contrast to using nod-averaged spectra for the continuum fitting. All the non-dereddened spectra are displayed in Fig.~\ref{spll} and the features listed in Table~\ref{raiecygx1}. Note that we only list the features detected at 5$\sigma$ or more above the continuum (assessed in 2~\mic\ windows using a second-order polynomial). 

As seen in Fig.~\ref{spll} and Table~\ref{raiecygx1}, the MIR spectra of \cygx1\ are dominated by \ion{H}{1} and \ion{He}{2} emission lines that likely arise from the stellar wind of HD~226868. Overall, more features are detected during Obs.~1 and Obs.~2, probably because of a higher soft X-ray emission flux level that photoionizes the wind. Moreover, in both cases, the spectroscopic content appears to vary from nod to nod, with more lines in nod~1 than in nod~2. This is likely due to slight background variations between both positions, although dramatic changes of the photoionization state cannot be excluded. 

During Obs.~3, the same features are detected at both positions with the exception of two absorption lines at $15.60\pm0.01$ and $25.90\pm0.02$~\mic\ in the nod~1 spectrum, which correspond to [\ion{Ne}{3}] and \ion{He}{2}. A forbidden absorption feature of \ion{Ne}{3} is very unlikely, and these two lines are probably a consequence of the sky subtraction. Indeed, given that slit positions differ between observations, the slit during Obs~3 may be aligned with non-homogeneous structures, leading to background variations between the two nods.

Finally, a total of three blue-shifted \ion{H}{1} emission lines are present in the nod~1 and nod~2 spectra of \cygx1\ during Obs.~2 but not during Obs.~1 nor Obs.~3. Their velocities vary between $-1600$ to $-2200$~km~s$^{-1}$, which hints at an origin from the stellar wind of HD~226868. This is likely due to the orbital configuration of the system during each observation. Indeed, \citet{2009Hanke} showed that one should expect the presence of blue- or red-shifted  X-ray absorption lines from the stellar wind, depending on the orbital phase of the BH (Fig. 14 in their paper). Close to phase 0.5, these lines are almost ``frozen'' while they are maximally blue-shifted around phases 0.2$-$0.3. During Obs.~1, Obs.~2, and Obs~3, the BH phases are 0.43, 0.27, and 0.58, respectively. Our results are therefore consistent with the prediction given in \citet{2009Hanke}.

\section{Discussion}

The main results of our multiwavelength study of \cygx1\ are: (1) the compact jets contribute significantly to the MIR emission, with a spectral break $\approx 2.9\times10^{13}$~Hz (Obs.~1), (2) when the jet carries more power, the cut-off frequency shifts as expected to higher frequencies and becomes undetectable in the MIR (Obs.~2), (3) the strength of the bremsstrahlung from the stellar winds is likely anti-correlated to the soft X-ray emission, which confirms  the anti-correlation between the mass-loss and accretion rates (all observations).

\subsection{Detection of the spectral break during Obs.~1}

This, to our knowledge, is the first measurement of the spectral break of a compact jet based on spectral fitting instead of photometry, and only the second detection of such a break in a BHXB after GX~339$-$4 \citep{2002Corbel}. We can use the theoretical expression of the expected cut-off frequency, to see if our value is consistent with that of GX~339$-$4, scaled with respect to the BH mass, distance, and accretion rate. In the HS, if $F_\nu$ is the flux density at a given frequency $\nu$, $\dot{M}$ the accretion rate, and $M_{\rm X}$ the BH mass, we have for a radiatively inefficient jet, from \citet{1995Falcke} and \citet{2003Heinz}: 
\begin{eqnarray}
\dot{M} &\propto& F_\nu^{12/17}\nonumber \\
\nu_{\rm b} &\propto& \frac{\dot{M}^{2/3}}{M_{\rm X}} 
\label{scale}
\end{eqnarray}
where $\nu_{\rm b}$ is the cut-off frequency. 

For GX~339$-$4, we consider a distance of 8~kpc, a BH mass of 10~\msun, and a flux density at 15~GHz of 15~mJy, respectively \citep{2004Zdzi,1997Fenderb, 2003Corbel}. For \cygx1\, the numbers are 2~kpc, 11~\msun\ \citep{2009Caballero}, and 15~mJy, respectively. Using Eq~\ref{scale}, we find $\nu_{\rm GX}\,\approx\,4.06\times\nu_{\rm CYG}$, where $\nu_{\rm GX}$ and $\nu_{\rm CYG}$ are the GX~339$-$4 and \cygx1\ cut-off frequencies, respectively. For our measured $\nu_{\rm CYG}\approx2.9\times10^{13}$~Hz, this relation gives $\nu_{\rm GX}\approx1.15\times10^{14}$~Hz, which is in good agreement with the values derived in \citet{2002Corbel}. Therefore, despite the intrinsic uncertainties associated with the MIR spectral fitting, we believe that we see the cut-off frequency in the \cygx1\ MIR spectrum during Obs.~1. 

\subsection{Implications for the jet geometry}

We can use the presence of the synchrotron break to constrain the jet geometry. As pointed out in \citet{2006Heinz}, the radio photosphere is located at a de-projected distance of about $5\times 10^{13}\,{\rm cm}$ from the black hole, based on the fact that about 50\% of the flux are unresolved at an angular resolution of 3 mas in the VLBA observations reported by \cite{2001Stirling}. 

For a Blandford-Koenigl jet \citep{1979Blandford}, the location of the photosphere along the jet, $z_{\tau=1}$ is inversely proportional to the frequency, $z_{\tau=1,\nu}\propto \nu^{-1}$.  Given the marginally higher radio flux in our observation compared to the Stirling et al. (2001) observations, the radio photosphere should be located at distance of about $6\times 10^{13}\,{\rm cm}$. At a frequency of $\nu_{\rm b}\approx 2.9\times 10^{13}\,{\rm Hz}$, the IR photosphere would be located at a distance of $z_{\tau=1,2.9\times 10^{13}\,{\rm Hz}} \approx 1.7\times 10^{10}\,{\rm cm}$. Since the break from optically thick to thin occurs when the photosphere reaches the base of the jet, our result would suggest that the base of the jet has a scale of about half a light second. For a BH of mass $M\approx 10\,M_{\odot}$, this corresponds to about 10,000 gravitational radii, which is fairly large compared to size scales typically assumed for the base of the jet, usually assumed to occur on scales of a few tens to hundreds of gravitational radii. For example, the jet in M87 is resolved down to scales of less than 100 gravitational radii.

A simple Blandford-Koenigl model is a severe oversimplification over the almost four orders of magnitude in frequency and thus photospheric scale between IR and radio. Deviations are almost certainly to be expected. However, any change would still have to obey the observed flat SED. As argued in \citet{2006Heinz}, simple geometric changes to the Blandford-Koenigl model do not alter the behavior of the jet significantly. For example, allowing the jet to be continually collimated (with $R_{\rm jet} \propto z^{\zeta}$) instead of being conical ($\zeta=1$) and requiring energy conservation such that $p \propto z^{-2\zeta}$ results in a slight altered dependence of $z_{\tau=1,\nu}\propto \nu^{-1/\zeta}$, and a change of the spectral index from completely flat to $\alpha = 1 - \frac{1}{\zeta}$ \citep{2006Heinz}. Bringing the base of the jet inwards (toward the BH) would require a $\zeta > 1$, implying a steeper spectrum, while microquasars tend to have flat to slightly inverted spectra. A simple geometric explanation for the large implied scale for the base of the jet is thus not plausible.

It is clear that the jet must be accelerated between the base of the jet and the radio photosphere, changing the Doppler parameter $\delta$ along the jet.  Since the optical depth to synchrotron-self absorption is proportional to $\delta^2$, a significant increase in $\delta$ from the base of the jet, where the IR emission originates, to the radio photosphere, would boost the radio optical depth, effectively moving the photosphere outward by a factor of $\delta^{2/3}$.  However, given that the jet in Cygnus X-1 is likely not ultra-relativistic \citep{2004Gleissner}, it is unlikely that this could bring the base of the jet inward by more than a factor of a few.

Finally, it is possible that the estimated resolved flux fraction of the radio of $\sim$ 50\% was overestimated in \citet{2006Heinz}. For example, if instead of 50\%, only 25\% of the flux were actually resolved in that observation, it would decrease $z_{\tau=0}$ by an order of magnitude, again moving the size of the IR photosphere and thus the base of the jet towards scales more typically associated with jet acceleration. However, very long baseline observations tend to underestimate the resolved flux, rather than the unresolved flux.

If the size scale over which the jet forms is really this large, it would imply that large scale outflows (i.e., winds) from the disk are fundamentally involved in jet formation. Indeed, a recent study of GRS~1915+105 have shown that the accretion disk wind and the jets were competing for the same matter supply \citep{2009Neilsen}. Moreover, it has long been speculated that winds support jet collimation by providing lateral pressure support. This would make the jet in Cygnus X-1 qualitatively different from AGN jets which have been imaged down to scales significantly smaller than this in terms of gravitational radii. It would also imply that the jet originates over a region much larger than the X-ray emitting corona that is typically assumed to be the base of the jet. This underscores the need for further confirmation of the nature of the outflow responsible for the observed radio emission and, by extension, the large scale diffuse radio and emission line nebula around \cygx1.

\subsection{Origins of the radio and X-ray/$\gamma$-ray emissions during Obs.~1}

\begin{figure*}
\begin{center}
\includegraphics[width=13cm]{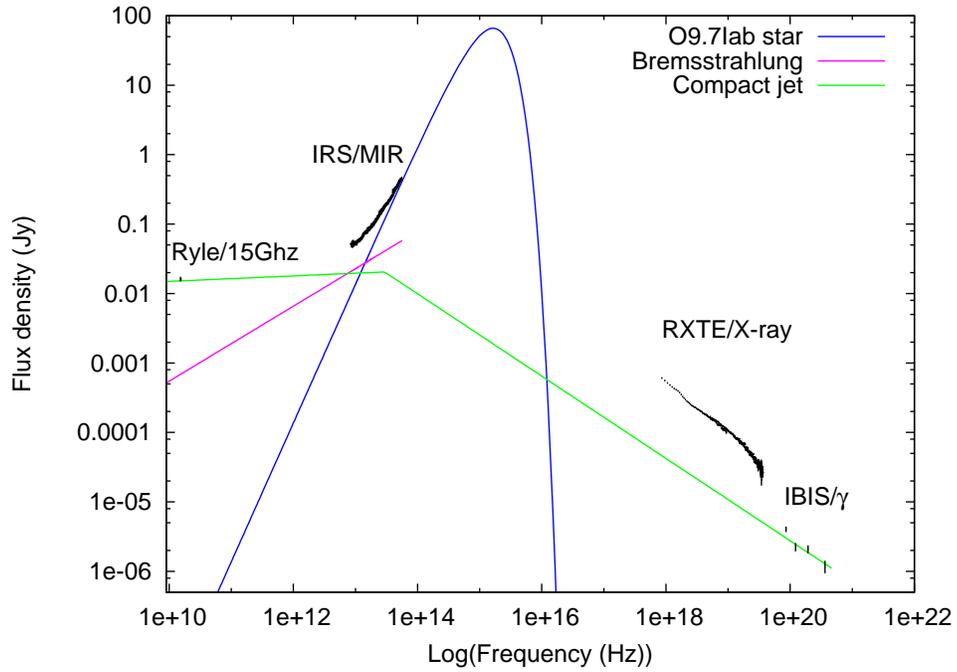}\\
\caption{\small Radio to X-ray SED of \cygx1\ during Obs.~1 (case 2a). The compact jet (green), bremsstrahlung (magenta), and stellar (blue) contributions are displayed. Simultaneous Ryle, IRS, and RXTE data are superimposed. We also add the data from the \textit{INTEGRAL}/IBIS compton mode, obtained by combining all the IBIS observations of the source, in the HS, between 2003 and 2010 \citep{2011Laurent}.}
\label{sedtot}
\end{center}
\end{figure*}

Fig~\ref{sedtot} displays the contributions of the stellar photosphere (blue), the bremsstrahlung from the expanding stellar wind (magenta), and the compact jet extrapolated to the high-energy (green), from our fit of the \cygx1's SED in the case 2, scenario (a). The simultaneous Ryle, \spitzer, and \rxte\ data are also superimposed, as well as the non-contemporaneous \textit{INTEGRAL}/IBIS data that were recently published \citep{2011Laurent}. The authors detected for the first time polarized $\gamma$-ray emission beyond 400~keV, and they argue that it is due to compact jets. Our result is fully consistent with this scheme, as the optically thin part of the jet perfectly describes the \textit{INTEGRAL}/IBIS data, providing strong evidence that we actually detect the spectral break around $2.9\times10^{13}$~Hz. Nevertheless,  it is clear that the \rxte\ spectrum of \cygx1\ cannot be described by the compact jet alone, and one or several extra components are needed. The spectral similarity between the spectral indices of the IR and the X-ray spectra might suggest that one of these components is synchrotron-self-compton (SSC) emission in the base of the compact jet itself \citep{2004Markoff, 2005Markoff}. However, this would require a large compactness of that region, in direct contradiction of our discussion regarding the implied large size of the IR photosphere and thus the base of the jet above. The more natural interpretation would be Comptonization of disk photons by the corona, independent of the optically thin synchrotron spectrum from the jet. Moreover, whether or not this result implies a disk origin of the X-ray emission in other sources like GX~339$-$4, whose NIR spectra are more consistent with a synchrotron origin of the X-ray, remains an open question.

\subsection{A clumpy stellar wind?}

Bremsstrahlung is found to be variable and to contribute significantly to the MIR emission of \cygx1, more than the compact jet itself.  The fact that its intensity is similar in the HS (Obs.~1 and Obs.~3), and lower in the failed-transition state (Obs.~2) is consistent with an anti-correlation of the mass-loss and accretion rates, as it was first shown by \citet{2003Gies}, who derived a mass-loss rate $\dot{M_{\rm w}}$ of about $2.57\times10^{-6}$~\msunyr\ and $2.00\times10^{-6}$~\msunyr\ for the HS and SS, respectively. We can also derive $\dot{M_{\rm w}}$ from the theoretical expression which links it to our fit-derived bremsstrahlung flux density $S_\nu$ at 15~GHz, including clumping effects \citep{1984Lamers, 1998Scuderi}: 

\begin{equation}
\dot{M_{\rm w}}=2.26\,10^{-7}\times\sqrt{f_\infty}\times\left[ S_{\nu} \left ( \frac{\nu}{10\,\textrm{GHz}} \right )^{-0.6} \left( \frac{T_{\rm e}}{10^{4}\,\textrm{K}} \right)^{-0.1} 
\left( \frac{D_\ast}{1\,\textrm{kpc}} \right )^2 \right ]^{\frac{3}{4}} \left ( \frac{\mu v_\infty}{100\,\textrm{km s}^{-1}} \right )
\label{eqbrem}
\end{equation}
where $f_\infty$ is the clump filling factor of the wind, $T_{\rm e}$ is the electron temperature far from the photosphere, which is well approximated as $0.5T_\ast$, $\mu$ is the mean atomic weight of the expanding plasma per electron, usually taken at about 1.3 for O supergiant \citep{1993Lamers}, and  $v_\infty$ is the terminal velocity of the wind \citep[$\approx 1600~\textrm{km s}^{-1}$][]{2008Gies}. When clumping is not included (i.e. $f_\infty=1$), we derive mass-loss rates of about $8.79\times10^{-6}$~\msunyr\ and $5.95\times10^{-6}$~\msunyr\ for Obs.~3 (HS) and Obs.~2 (IS), respectively. Neglecting clumping clearly leads to overestimate the mass-loss rate by a factor of 3$-$4, as shown by \citet{1984Lamers}. This is therefore a hint that the \cygx1\ wind is clumped, and we need $f_\infty\approx0.09-0.10$ in Eq.~\ref{eqbrem} to match the mass-loss rates given in \citet{2008Gies}. This value is consistent with that derived from the study of the X-ray dips (Hanke et al., in prep).

 \section{Conclusion}

We have investigated, through SED fitting, the origin of the MIR emission of \cygx1\ when the source was in the HS (Obs.~1), the IS (Obs.~2), and in a peculiar compact jet-free HS (Obs.~3). We interpret the MIR continuum during Obs.~3 as due to the stellar's photosphere and clumpy expanding wind ($f_\infty\approx0.09-0.10$). In contrast, when  jets are present, they clearly contribute to the MIR emission. During Obs.~1, we detect the spectral break ($\nu_{\rm b}\approx2.9\times10^{13}$~Hz), and we argue that this value is consistent with the base of the jet being very large, likely involving accretion disk wind in the formation and collimation of the jet. We show that if the jet alone can account for the $\gamma$-ray emission of the source, additional processes are needed to explain the $3-200$~keV spectrum. A compact jet is also present during Obs.~2 but we do not detect the cut-off frequency as it has likely shifted towards higher frequencies. Nonetheless, the bremsstrahlung emission is found lower than that during Obs.~3,  which is a confirmation of the anti-correlation between the mass-loss and accretion rates. 

Our results highlight the importance of the MIR and, by extension, the far-infrared in the study of microquasars, as all their components may have a spectral signature in these domains. The existence of powerful infrared space-based observatories, \textit{Herschel} today and the James Webb Space Telescope (\textit{JWST}) in the near future, combined to ground-based telescope such as the Atacama Millimeter/sub-millimeter Array (ALMA), are therefore a unique opportunity to make strong progresses in our understanding of their behavior and properties. We definitively recommend further studies of \cygx1\ and other microquasars with these facilities.

\acknowledgments
We thank the anonymous referee for his/her useful comments. FR thanks J. 
Rodriguez for providing the \textit{INTEGRAL}/IBIS data. JCL thanks the Harvard Faculty of Arts and Sciences and the Harvard College Observatory. SH acknowledges support from  JPL-NASA contract No.~1292543. This work was partially funded by the Bundesministerium f\"ur Wirtschaft und Technologie through Deutsches Zentrum f\"ur Luft- und Raumfahrt grant 50\,OR\,1007 and by
the European Commission through contract ITN~215212 ``Black Hole Universe''.
This research has made use of NASA's Astrophysics Data System, of the SIMBAD, and VizieR databases operated at CDS, 
Strasbourg, France.
\bibliographystyle{apj}
\bibliography{./mybib}{}

\begin{deluxetable}{cccccc}
\tabletypesize{\scriptsize}
\rotate
\tablecaption{\small Summary of all the observations of \cygx1\ we made use of in this study\label{logobs}}
\tablewidth{0pt}
\tablehead{\colhead{Obs.~\#}&\colhead{IRS}&\colhead{\rxte}&\colhead{Ryle}&\colhead{Ryle fluxes (mJy)}&\colhead{$\Phi$}}
\startdata
1&53513.0399839$-$53513.0544390&53513.0433283$-$53513.2142542&53513.0398560$-$53513.0546265&5$-$28&0.426\\
2&53528.9592175$-$53528.9736728&53528.9698098$-$53529.2433283&53528.9593506$-$53528.9737549&4$-$42&0.269\\
3&53553.1129871$-$53553.1274610&53552.9079579$-$53553.1988838&53553.1131592$-$53553.1275635&0$-$48&0.583\\
\enddata
\tablecomments{We give the observation number, the IRS day of observation (in MJD), the day of simultaneous coverage with \rxte\ and the Ryle telescope, the Ryle flux level at 15~GHz (in mJy), as well as the orbital phase $\Phi$ calculated from the ephemeris given in \citet{1999brocksoppa}.} 
\end{deluxetable}

\begin{deluxetable}{cccccccccccc}
  \tablewidth{0pt}
  \tabletypesize{\scriptsize}
  \rotate
  \tablecaption{\small Best parameters from the fits of Obs.~1, Obs.~2, and Obs.~3 \rxte/PCA+HEXTE spectra of \cygx1 \label{xpar}}
  \tablehead{\colhead{Obs. \#}&\colhead{$N_{\rm H}$}&\colhead{$E_{\rm cut}$}&\colhead{$E_{\rm fold}$}&\colhead{$\Gamma_1$}&\colhead{$E_{\rm break}$}&\colhead{$\Gamma_2$}&\colhead{$E_{\rm iron}$}&\colhead{$\sigma_{\rm iron}$}&\colhead{$F_{\rm 3-20}$}&\colhead{$F_{\rm 20-200}$}&\colhead{\chis2/dof}\\
 \colhead{\nodata}&\colhead{\tiny ($10^{22}$~\cm2)}&\colhead{\tiny (keV)}&\colhead{\tiny (keV)}&\colhead{\nodata}&\colhead{\tiny (keV)}&\colhead{\nodata}&\colhead{\tiny (keV)}&\colhead{\tiny (keV)}&\colhead{\tiny $(10^{-8}$\ergcms$)$}&\colhead{\tiny $(10^{-8}$\ergcms$)$}&\colhead{\nodata}} 
 \startdata
  1&$0.68_{-0.18}^{+0.16}$&$26.15_{-1.89}^{+2.09}$&$175.97_{-6.70}^{+6.97}$&$1.83_{-0.01}^{+0.01}$&$10.24_{-0.26}^{+0.25}$&$1.56_{-0.01}^{+0.01}$&$6.38_{-0.06}^{+0.06}$&$0.60_{-0.08}^{+0.08}$&1.19&2.41&1.02/303\\
  2&0.53 (frozen)&$26.87_{-2.09}^{+2.12}$&$151.67_{-6.78}^{+7.35}$&$2.08_{-0.01}^{+0.01}$&$10.33_{-0.19}^{+0.18}$&$1.70_{-0.01}^{+0.01}$&$6.45_{-0.03}^{+0.03}$&$0.58_{-0.05}^{+0.05}$&1.55&2.21&1.08/304\\
  3&$1.04_{-0.11}^{+0.11}$&$25.28_{-3.82}^{+2.60}$&$212.18_{-9.09}^{+9.56}$&$1.75_{-0.01}^{+0.01}$&$10.11_{-0.23}^{+0.26}$&$1.53_{-0.01}^{+0.01}$&$6.41_{-0.04}^{+0.04}$&$0.39_{-0.07}^{+0.07}$&0.71&1.64&1.09/303\\
 \enddata
  \tablecomments{The best-fit model is \textit{phabs(highecut$\times$bknpower+gaussian)}, and the errorbars are given at the 90\% confidence level.}
  \tablenotetext{a}{The fluxed are unabsorbed and exclude that of the iron feature}
\end{deluxetable}

\begin{deluxetable}{cccccc}
  \tablewidth{0pt}
  \tabletypesize{\footnotesize}
  \rotate
  \tablecaption{\small Best parameters from the fits of the Obs.~3 MIR spectrum of \cygx1 \label{mirparobs3}}
  \tablehead{\colhead{Obs. \#}&\colhead{$T_\ast$}&\colhead{$R_{\ast}/D_{\ast}$}&\colhead{$\alpha_{\rm ff}$}&\colhead{$N_{\rm ff}$}&\colhead{\chis2/dof}\\
 \colhead{\nodata}&\colhead{\tiny (K)}&\colhead{\tiny ($R_\odot$/kpc)}&\colhead{\nodata}&\colhead{\tiny (mJy at 15~GHz)}&\colhead{\nodata}} 
 \startdata
  3&28000 (frozen)&$9.94_{-0.03}^{+0.03}$&$0.55_{-0.07}^{+0.07}$&$0.74_{-0.10}^{+0.07}$&1.00/302\\
\enddata
  \tablecomments{The model used is a combination of a black body and a power law. The errorbars are given at the 90\% confidence level.}
\end{deluxetable}

\begin{deluxetable}{cccccccc}
  \tablewidth{0pt}
  \tabletypesize{\footnotesize}
  \rotate
  \tablecaption{\small Best parameters from the fits of Obs.~1 and Obs.~2 radio/MIR SEDs of \cygx1: case 1 \label{mirparcase1}}
  \tablehead{\colhead{Obs. \#}&\colhead{$T_\ast$}&\colhead{$R_{\ast}/D_{\ast}$}&\colhead{$\alpha_{\rm ff}$}&\colhead{$N_{\rm ff}$}&\colhead{$\alpha_{\rm j}$}&\colhead{$N_{\rm j}$}&\colhead{\chis2/dof}\\
 \colhead{\nodata}&\colhead{\tiny (K)}&\colhead{\tiny ($R_\odot$/kpc)}&\colhead{\nodata}&\colhead{\tiny (mJy at 15~GHz)}&\colhead{\nodata}&\colhead{\tiny (mJy at 15~GHz)}&\colhead{\nodata}} 
 \startdata
 1&28000~(\textrm{frozen})&9.94~(\textrm{frozen})&0.55~(\textrm{frozen})&$0.66_{-0.05}^{+0.05}$&$0.01_{-0.02}^{+0.02}$&$15.60_{-4.24}^{+5.41}$&0.96/301\\
 2&=&=&=&$0.44_{-0.05}^{+0.06}$&$0.06_{-0.02}^{+0.02}$&$22.00_{-4.32}^{+5.28}$&0.72/301\\
 \enddata
  \tablecomments{The model used is a combination of a black body and two power laws. The errorbars are given at the 90\% confidence level.}
\end{deluxetable}

\begin{deluxetable}{cccccccccc}
  \tablewidth{0pt}
  \tabletypesize{\footnotesize}
  \rotate
  \tablecaption{\small Best parameters from the fits of Obs.~1 and Obs.~2 radio/MIR SEDs of \cygx1: case 2, scenario (a) \label{mirparcase21}}
  \tablehead{\colhead{Obs. \#}&\colhead{$T_\ast$}&\colhead{$R_{\ast}/D_{\ast}$}&\colhead{$\alpha_{\rm ff}$}&\colhead{$N_{\rm ff}$}&\colhead{$\alpha_{\rm j1}$}&\colhead{$\nu_{\rm c}$}&\colhead{$\alpha_{\rm j2}$}&\colhead{$N_{\rm j}$}&\colhead{\chis2/dof}\\
 \colhead{\nodata}&\colhead{\tiny (K)}&\colhead{\tiny ($R_\odot$/kpc)}&\colhead{\nodata}&\colhead{\tiny (mJy at 15~GHz)}&\colhead{\nodata}&\colhead{\tiny (THz)}&\colhead{\nodata}&\colhead{\tiny (mJy at 15~GHz)}&\colhead{\nodata}} 
 \startdata
 1&28000~(\textrm{frozen})&9.94~(\textrm{frozen})&0.55~(\textrm{frozen})&$0.75_{-0.04}^{+0.03}$&$0.01_{-0.02}^{+0.02}$&$27.05_{-1.78}^{+2.44}$&$-0.6$~(frozen)&$15.40_{-3.34}^{+4.43}$&0.81/300\\
 2&=&=&=&$0.78_{-0.04}^{+0.02}$&$-0.01_{-0.01}^{+0.01}$&$14.80_{-1.07}^{+1.16}$&$-0.6$~(frozen)&$21.30_{-2.45}^{+3.39}$&0.69/300\\
 \enddata
   \tablecomments{The model used is a combination of a black body, a power law, and a broken power law. The errorbars are given at the 90\% confidence level.}
\end{deluxetable}

\begin{deluxetable}{cccccccccc}
  \tablewidth{0pt}
  \tabletypesize{\footnotesize}
  \rotate
  \tablecaption{\small Best parameters from the fits of Obs.~1 and Obs.~2 radio/MIR SEDs of \cygx1: case 2, scenario (b) \label{mirparcase22}}
  \tablehead{\colhead{Obs. \#}&\colhead{$T_\ast$}&\colhead{$R_{\ast}/D_{\ast}$}&\colhead{$\alpha_{\rm ff}$}&\colhead{$N_{\rm ff}$}&\colhead{$\alpha_{\rm j1}$}&\colhead{$\nu_{\rm c}$}&\colhead{$\alpha_{\rm j2}$}&\colhead{$N_{\rm j}$}&\colhead{\chis2/dof}\\
 \colhead{\nodata}&\colhead{\tiny (K)}&\colhead{\tiny ($R_\odot$/kpc)}&\colhead{\nodata}&\colhead{\tiny (mJy at 15~GHz)}&\colhead{\nodata}&\colhead{\tiny (THz)}&\colhead{\nodata}&\colhead{\tiny (mJy at 15~GHz)}&\colhead{\nodata}} 
 \startdata
 1&28000~(\textrm{frozen})&9.94~(\textrm{frozen})&0.55~(\textrm{frozen})&0.66~(frozen)&$0.04_{-0.01}^{+0.01}$&$29.43_{-2.29}^{+2.38}$&$-0.58_{-0.21}^{+0.27}$&$15.20_{-0.09}^{+0.11}$&0.85/300\\
 2&=&=&=&0.44~(frozen)&$0.06_{-0.01}^{+0.01}$&$42.61_{-1.79}^{+3.73}$&$1.21_{-0.36}^{+0.80}$&$22.04_{-0.27}^{+0.28}$&0.69/300\\
 \enddata
  \tablecomments{The model used is a combination of a black body, a power law, and a broken power law. The errorbars are given at the 90\% confidence level.}
\end{deluxetable}

\begin{deluxetable}{cccccccccc}
  \tablewidth{0pt}
  \tabletypesize{\footnotesize}
  \rotate
  \tablecaption{\small Best parameters from the fits of Obs.~1 and  Obs.~2 radio/MIR SEDs of \cygx1: case 2, scenario (c) \label{mirparcase23}}
  \tablehead{\colhead{Obs. \#}&\colhead{$T_\ast$}&\colhead{$R_{\ast}/D_{\ast}$}&\colhead{$\alpha_{\rm ff}$}&\colhead{$N_{\rm ff}$}&\colhead{$\alpha_{\rm j1}$}&\colhead{$\nu_{\rm c}$}&\colhead{$\alpha_{\rm j2}$}&\colhead{$N_{\rm j}$}&\colhead{\chis2/dof}\\
 \colhead{\nodata}&\colhead{\tiny (K)}&\colhead{\tiny ($R_\odot$/kpc)}&\colhead{\nodata}&\colhead{\tiny (mJy at 15~GHz)}&\colhead{\nodata}&\colhead{\tiny (THz)}&\colhead{\nodata}&\colhead{\tiny (mJy at 15~GHz)}&\colhead{\nodata}} 
 \startdata
 1&28000 (frozen)&9.94 (frozen)&0.55 (frozen)&0.74 (frozen)&$0.01_{-0.01}^{+0.01}$&$29.27_{-1.91}^{+1.90}$&$-1.14_{-0.34}^{+0.44}$&$15.30_{-1.41}^{+1.65}$&0.78/300\\
 2&=&=&=&=&$0.01_{-0.01}^{+0.01}$&$15.28_{-1.16}^{+1.24}$&$-0.55_{-0.10}^{+0.04}$&$21.60_{-2.30}^{+2.80}$&0.69/300\\
 \enddata
  \tablecomments{The model used is a combination of a black body, a power law, and a broken power law. The errorbars are given at the 90\% confidence level.}
\end{deluxetable}

\newpage
\begin{deluxetable}{c|cccccc||cccccc}
\rotate
\tabletypesize{\tiny}
\tablecaption{\small List of all detected features for each LL nodding position \label{raiecygx1}}
\tablewidth{0pt}
\tablehead{&\multicolumn{6}{c||}{Nod 1}&\multicolumn{6}{c}{Nod 2}\\
\hline
&Feature&$\lambda_{\rm fit}$&$\lambda$&$\mathring{W}$&\textit{FWHM}&Flux&Feature&$\lambda_{\rm fit}$&$\lambda$&$\mathring{W}$&\textit{FWHM}&Flux\\
&\nodata&\tiny (\mic)&\tiny (\mic)&\tiny (\mic)&\tiny (\mic)&\tiny ($\times 10^{-21}$~W~cm$^{-2}$)&\nodata&\tiny (\mic)&\tiny (\mic)&\tiny (\mic)&\tiny (\mic)&\tiny ($\times 10^{-21}$~W~cm$^{-2}$)}
\startdata
\multirow{8}{*}{Obs.~1}&\iontiny{H}{i}&14.68$\pm$0.01&14.71&-0.010$\pm$0.002&0.21$\pm$0.01&1.52$\pm$0.29&&&&&&\\ 
&\iontiny{H}{i}&15.83$\pm$0.01&15.82&-0.006$\pm$0.002&0.13$\pm$0.01&0.48$\pm$0.13&&&&&&\\
&&&&&&&\iontiny{He}{ii}&16.15$\pm$0.01&16.20&-0.008$\pm$0.001&0.24$\pm$0.01&0.96$\pm$0.16\\
&\iontiny{H}{i}&16.89$\pm$0.02&16.88&-0.018$\pm$0.003&0.35$\pm$0.02&1.48$\pm$0.26&\iontiny{H}{i}&16.88$\pm$0.02&16.88&-0.019$\pm$0.004&0.37$\pm$0.02&1.22$\pm$0.23\\
&\iontiny{H}{i}&18.01$\pm$0.02&18.04&-0.018$\pm$0.004&0.26$\pm$0.02&1.17$\pm$0.28&&&&&&\\
&\iontiny{H}{i}&18.62$\pm$0.02&18.62&-0.027$\pm$0.005&0.29$\pm$0.02&1.64$\pm$0.29&\iontiny{H}{i}&18.64$\pm$0.01&18.62&-0.022$\pm$0.001&0.28$\pm$0.01&1.29$\pm$0.11\\
&\iontiny{He}{ii}&19.08$\pm$0.02&19.10&-0.016$\pm$0.006&0.17$\pm$0.02&0.92$\pm$0.32&\iontiny{He}{ii}&19.10$\pm$0.01&19.10&-0.015$\pm$0.002&0.23$\pm$0.01&0.86$\pm$0.06\\
&\iontiny{He}{ii}&22.33$\pm$0.01&22.33&-0.021$\pm$0.002&0.43$\pm$0.01&0.70$\pm$0.06&&&&&&\\
\hline
\multirow{9}{*}{Obs.~2}&\iontiny{H}{i}&14.71$\pm$0.01&14.71&-0.006$\pm$0.001&0.16$\pm$0.01&0.65$\pm$0.14&&&&&&\\ 
&&&&&&&\iontiny{H}{i}$^\dagger$&14.88$\pm$0.03&14.96&-0.003$\pm$0.001&0.11$\pm$0.01&0.33$\pm$0.08\\
&\iontiny{H}{i}&15.20$\pm$0.01&15.20&-0.006$\pm$0.001&0.15$\pm$0.01&0.60$\pm$0.12&&&&&&\\
&\iontiny{H}{i}$^\dagger$&17.87$\pm$0.02&18.04&-0.007$\pm$0.002&0.18$\pm$0.02&0.48$\pm$0.13&&&&&&\\
&\iontiny{H}{i}$^\dagger$&18.50$\pm$0.02&18.62&-0.010$\pm$0.002&0.23$\pm$0.02&0.63$\pm$0.12&&&&&&\\
&\iontiny{He}{ii}&19.10$\pm$0.01&19.10&-0.007$\pm$0.003&0.11$\pm$0.02&0.47$\pm$0.21&\iontiny{He}{ii} &19.10$\pm$0.02&19.10&-0.010$\pm$0.006&0.07$\pm$0.02&0.47$\pm$0.30\\
&\iontiny{He}{ii}&20.04$\pm$0.01&20.05&-0.010$\pm$0.002&0.13$\pm$0.01&0.46$\pm$0.09&&&&&&\\
&\iontiny{He}{ii}&22.18$\pm$0.02&22.18&-0.017$\pm$0.005&0.20$\pm$0.03&0.54$\pm$0.20&&&&&&\\
\hline
\multirow{5}{*}{Obs.~3}&\iontiny{H}{i}&14.69$\pm$0.01&14.71&-0.012$\pm$0.002&0.14$\pm$0.01&1.22$\pm$0.11&\iontiny{H}{i}&14.71$\pm$0.01&14.71&-0.011$\pm$0.001&0.26$\pm$0.01&1.31$\pm$0.09\\ 
&$[$\iontiny{Ne}{iii}$]$&15.60$\pm$0.01&15.55&0.007$\pm$0.003&0.17$\pm$0.01&0.92$\pm$0.19&&&&&&\\
&\iontiny{He}{ii}&16.21$\pm$0.01&16.20&-0.012$\pm$0.001&0.12$\pm$0.09&0.84$\pm$0.07&\iontiny{He}{ii}&16.21$\pm$0.01&16.20&-0.008$\pm$0.003&0.32$\pm$0.02&0.64$\pm$0.16\\
&\iontiny{He}{ii}&19.09$\pm$0.01&19.10&-0.019$\pm$0.002&0.26$\pm$0.06&1.19$\pm$0.07&\iontiny{He}{ii} &19.12$\pm$0.02&19.10&-0.041$\pm$0.001&0.18$\pm$0.02&1.74$\pm$0.06\\
&\iontiny{He}{ii}&25.90$\pm$0.02&25.90&0.082$\pm$0.001&0.26$\pm$0.02&1.53$\pm$0.04&&&&&&\\
\enddata
\tablecomments{We give the name of the feature, its measured ($\lambda_{\rm fit}$) and laboratory ($\lambda$) wavelengths, 
its equivalent width $\mathring{W}$, its full-width at half-length ($FWHM$), and its flux.\\
$^\dagger$ Blue-shifted lines}

\end{deluxetable}

\end{document}